\documentclass[lettersize,journal]{IEEEtran}
\IEEEoverridecommandlockouts

\usepackage{cite}
\usepackage{amsmath,amssymb,amsfonts}
\usepackage{algorithmic}
\usepackage{graphicx}
\usepackage{textcomp}
\usepackage{xcolor}
\def\BibTeX{{\rm B\kern-.05em{\sc i\kern-.025em b}\kern-.08em
    T\kern-.1667em\lower.7ex\hbox{E}\kern-.125emX}}
\usepackage{multirow}
\usepackage{algorithm}
\usepackage{array}
\usepackage{geometry}
\usepackage[caption=false,font=normalsize,labelfont=sf,textfont=sf]{subfig}
\usepackage{textcomp}
\usepackage{stfloats}
\usepackage{url}
\usepackage{xcolor}
\usepackage{verbatim}
\usepackage{caption} 
\usepackage{graphicx}
\usepackage{subfig}
\usepackage{enumerate}
\usepackage{enumitem}
\usepackage{tabularx,booktabs,textcomp}
\usepackage{amsthm}

\usepackage{amsmath,amssymb,amsthm}

\makeatletter 
\newcommand{\linebreakand}{%
  \end{@IEEEauthorhalign}
  \hfill\mbox{}\par
  \mbox{}\hfill\begin{@IEEEauthorhalign}
}

\makeatother

\begin{document}

\title{Efficient Fault Detection and Categorization in Electrical Distribution Systems Using Hessian Locally Linear Embedding on Measurement Data}

\author{{K. Victor Sam Moses Babu, Sidharthenee Nayak, Divyanshi Dwivedi, Pratyush Chakraborty, Chandrashekhar
Narayan Bhende,  Pradeep Kumar Yemula, Mayukha Pal}
 

\thanks{(Corresponding author: Mayukha Pal)}

\thanks{Mr. K. Victor Sam Moses Babu is a Data Science Research Intern at ABB Ability Innovation Center, Hyderabad 500084, India and also a Research Scholar at the Department of Electrical and Electronics Engineering, BITS Pilani Hyderabad Campus, Hyderabad 500078, IN.}
\thanks{Ms. Sidharthenee Nayak is a Data Science Research Intern at ABB Ability Innovation Center, Hyderabad 500084, India, and also a postgraduate student at the School of Electrical Sciences, Indian Institute of Technology, Bhubaneswar, 751013, IN.}
\thanks{Mrs. Divyanshi Dwivedi is a Data Science Research Intern at ABB Ability Innovation Center, Hyderabad 500084, India, and also a Research Scholar at the Department of Electrical Engineering, Indian Institute of Technology, Hyderabad 502205, IN.}
\thanks{Dr. Pratyush Chakraborty is an Asst. Professor with the Department of Electrical and Electronics Engineering, BITS Pilani Hyderabad Campus, Hyderabad 500078, IN.}
\thanks{Dr. Chandrashekhar Narayan Bhende is a Professor at the School of Electrical Sciences
Indian Institute of Technology
Bhubaneswar, 751013, IN.}
\thanks{Dr. Pradeep Kumar Yemula is an Assoc. Professor with the Department of Electrical Engineering, Indian Institute of Technology, Hyderabad 502205, IN.}
\thanks{Dr. Mayukha Pal is with ABB Ability Innovation Center, Hyderabad-500084, IN, working as Global R\&D Leader – Cloud \& Analytics (e-mail: mayukha.pal@in.abb.com).}
}

\maketitle

\begin{abstract}
Faults on electrical power lines could severely compromise both the reliability and safety of power systems, leading to unstable power delivery and increased outage risks. They pose significant safety hazards, necessitating swift detection and mitigation to maintain electrical infrastructure integrity and ensure continuous power supply. Hence, accurate detection and categorization of electrical faults are pivotal for optimized power system maintenance and operation. In this work, we propose a novel approach for detecting and categorizing electrical faults using the Hessian locally linear embedding (HLLE) technique and subsequent clustering with t-SNE (t-distributed stochastic neighbor embedding) and Gaussian mixture model (GMM). First, we employ HLLE to transform high-dimensional (HD) electrical data into low-dimensional (LD) embedding coordinates. This technique effectively captures the inherent variations and patterns in the data, enabling robust feature extraction. Next, we perform the Mann-Whitney U test based on the feature space of the embedding coordinates for fault detection. This statistical approach allows us to detect electrical faults providing an efficient means of system monitoring and control. Furthermore, to enhance fault categorization, we employ t-SNE with GMM to cluster the detected faults into various categories. To evaluate the performance of the proposed method, we conduct extensive simulations on an electrical system integrated with solar farm. Our results demonstrate that the proposed approach exhibits effective fault detection and clustering across a range of fault types with different variations of the same fault. Overall, this research presents an effective methodology for robust fault detection and categorization in electrical systems, contributing to the advancement of fault management practices and the prevention of system failures.
\end{abstract}

\begin{IEEEkeywords}
Electrical fault detection, fault clustering, Hessian locally linear embedding, Gaussian mixture model, Mann-Whitney U test,t-distributed stochastic neighbor embedding.
\end{IEEEkeywords}

\section{Introduction}
\label{sec:Introduction}

The electrical power grid serves as the backbone of modern society by consistently delivering electricity to homes, businesses, and industries. With our increasing reliance on electricity, the demand for strong and efficient electrical distribution systems (EDSs) continues to rise. In order to ensure the reliability and safety of these systems, it is essential to minimize potential risks and maintain a continuous power supply. To enhance the operation of EDSs, it's imperative to implement advanced fault detection and categorization techniques \cite{i1}. These techniques serve as a critical tool in identifying and managing faults, optimizing maintenance efforts, and ultimately strengthening the overall resilience of the grid \cite{DWIVEDI2023156}.

Fault detection and clustering (FDC) plays a crucial role in minimizing the restoration costs associated with distribution lines and enhancing the overall protection of the power network. When faults occur in distribution lines, they disrupt the power supply of end users. Various types of faults, such as double line (LL), single line-to-ground (SLG), double line-to-ground (DLG), and triple line-to-ground faults (TLG), etc., could occur in EDS. Among these faults, SLG faults are the most frequent (70\%) but generally less severe, while three-phase faults are the least frequent (5\%) but are the most severe \cite{STEFANIDOUVOZIKI2022108031}. These faults may cause overheating of components, mechanical stress, and current imbalance within the system \cite{i2}. The restoration of faulty phases often relies on human intervention; it depends on the detection and categorization of faults to determine their types within the system. Fast isolation and removal of faulty phases from the network is possible because of fast and accurate FDC. Removing these faulty phases improves power quality, enhances system stability, and increases transient stability in the power network \cite{i3,i4}.

Numerous prior research efforts have been dedicated to detecting and categorizing the faults, which include classical methods based on basic circuit theory \cite{cl1,cl2,cl3}, machine learning (ML) models \cite{ml1,ml2,ml3,ml4, ml5, ml6}, transform-based feature extraction models \cite{dwt1,dwt2,w1, wt4}, fuzzy methods \cite{fs1}, and expert system-based models \cite{es1}. Classical methods are known for their simplicity in both calculation and implementation. To support the execution, protective devices have been installed to expedite decision-making. These protective devices rely on synchronized measurements, indicating the difficulty of dealing with timing issues in the absence of phasor measurement units \cite{cl2,cl3}. Also, classical methods may yield inaccurate results due to assumptions about distribution line parameters \cite{cl1}.

As EDSs are becoming more complex \cite{Victor_EVCI}, FDC tasks require artificial neural network-based algorithms capable of handling large volumes of data, including feed-forward functions \cite{SALSBURY2001403}, probabilistic functions \cite{25}, and radial basis functions \cite{rbf}. These models aim to develop intelligent systems for fault detection and categorization and 
react to new scenarios using historical data from faulted voltage and current signals across various phases of the distribution lines \cite{ml1}. In \cite{27}, a feedback neural network (FNN) is used for identifying tree and animal contact disruptions in EDS. The radial-basis neural network (RBNN) improves upon FNN with a sigmoid activation function, achieving better fault identification results using radial basis functions \cite{rbf1}. In contrast, the probabilistic neural network combines discrete wavelet transform (DWT) and exponential activation function to categorize faults with 10\% improvement in accuracy than the probabilistic neural network \cite{pnn}. However, they face generalization issues because the model performance and accuracy are highly related to the selection of features.

Recently, machine learning algorithms have been widely used by combining signal processing approaches to rapidly and accurately identify faults \cite{w1,ml3,ml5}. Signal processing techniques extract the features from initially captured voltage and current signals to determine fault occurrence and their types \cite{5}. However, categorizing faults based solely on raw three-phase data would be challenging. Consequently, various tools like wavelet transform \cite{dwt1,dwt2, wt4}, Hilbert-Huang transforms \cite{cl3}, Clarke transforms \cite{ct1}, and S-transforms \cite{st1} are used to extract fault features from line signals, enabling the characterization of fault nature. Unfortunately, the selection of faulted features across different frequency ranges is often arbitrary, leading to inconsistency in results. Therefore, enhancing the FDC's accuracy for EDSs has emerged as a significant research focus. Also, existing FDC techniques are supervised approach, which poses challenges for real-time applications due to the requirement of prior labeling, and achieving online fault detection and clustering with a high degree of accuracy remains elusive.

Our proposed approach focuses on addressing these challenges through the analysis of HD measurement data with an emphasis on simplicity and efficiency. The general process flow for fault detection using feature extraction is shown in Fig. \ref{fig:Basic_process}. We employ an online fault detection approach based on the Hessian locally linear embedding technique as described in \cite{hlle2}. HLLE is an effective method that enables efficient analysis and reliable information while preserving the intricate nonlinear relationships among various components \cite{hlle3, hlle4}. In \cite{hlle1}, HLLE with $T^2$ statistical test is used for fault detection in PMU data, but the method is tested for only two fault cases in the transmission system. For the considered PMU data, the $T^2$ statistic was found to be suitable. In comparison to this work, we have simulated a large dataset of 10 various types of faults with varying fault resistances and locations in a distribution system with distributed energy resources (DER) integration. We have considered the detection of faults in continuous streaming data by using the Mann-Whitney U statistical test between a reference segment and each subsequent streaming segment of data. We then account for the values obtained from these tests in a time series format. A fault is detected when the value significantly deviates from the reference value. This approach offers both visual and quantitative insights into the system's state evolution over time. Using the obtained HLLE values and raw streaming data, we perform unsupervised clustering using the Gaussian mixture model to categorize the faults without having prior knowledge about the fault event. This work offers a comprehensive and innovative solution to the critical problem of electrical fault detection and categorization, leveraging advanced techniques to improve the efficiency and reliability of power system operation and maintenance. The significant contributions of this work are as follows:

\begin{figure}[t]
  \centering
  \includegraphics[width=3.4in]{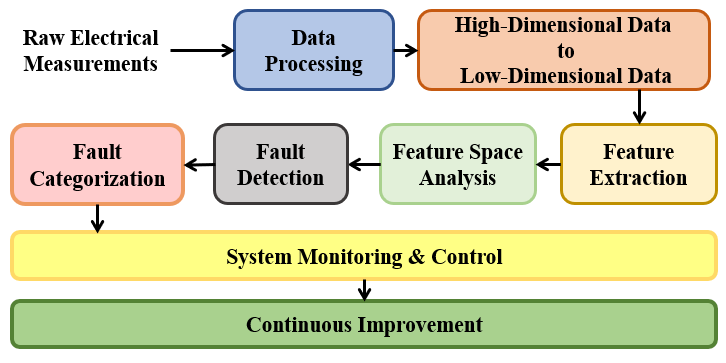}
  \caption{Block diagram of the general process flow for fault detection using feature extraction.}
  \label{fig:Basic_process}
\end{figure}

\begin{enumerate}
    \item A novel online approach for detecting and categorizing electrical faults, combining advanced data analysis techniques, is proposed. The proposed methodology is tested for large simulated data with 10 different fault types considering several variations of each fault.
    \item The use of Hessian locally linear embedding to transform HD electrical data into LD embedding coordinates captures complex variations and patterns in the data, facilitating robust feature extraction, which is vital for accurate fault detection. We use the Mann-Whitney U test on the feature space derived from the embedding coordinates to detect electrical faults efficiently. This statistical technique proves effective in detecting various types of electrical faults, thereby enhancing system monitoring and control capabilities.
    \item For fault categorization, we integrate t-SNE with a Gaussian mixture model. This integration enables the clustering of detected faults into various categories in an unsupervised manner, eliminating the need for prior labeling and streamlining the fault categorization process.
\end{enumerate}

The remainder of the paper is structured as follows: section \ref{sec:Method} presents a detailed description of our methodology, including the HLLE, Mann-Whitney U test, and t-SNE with GMM processes. In section \ref{sec:results}, comprehensive information on the simulation model, results analysis, and discussions are detailed. Lastly, section \ref{sec:conclusion} concludes the paper.

\begin{figure*}
  \centering
  \includegraphics[width=7.4in]{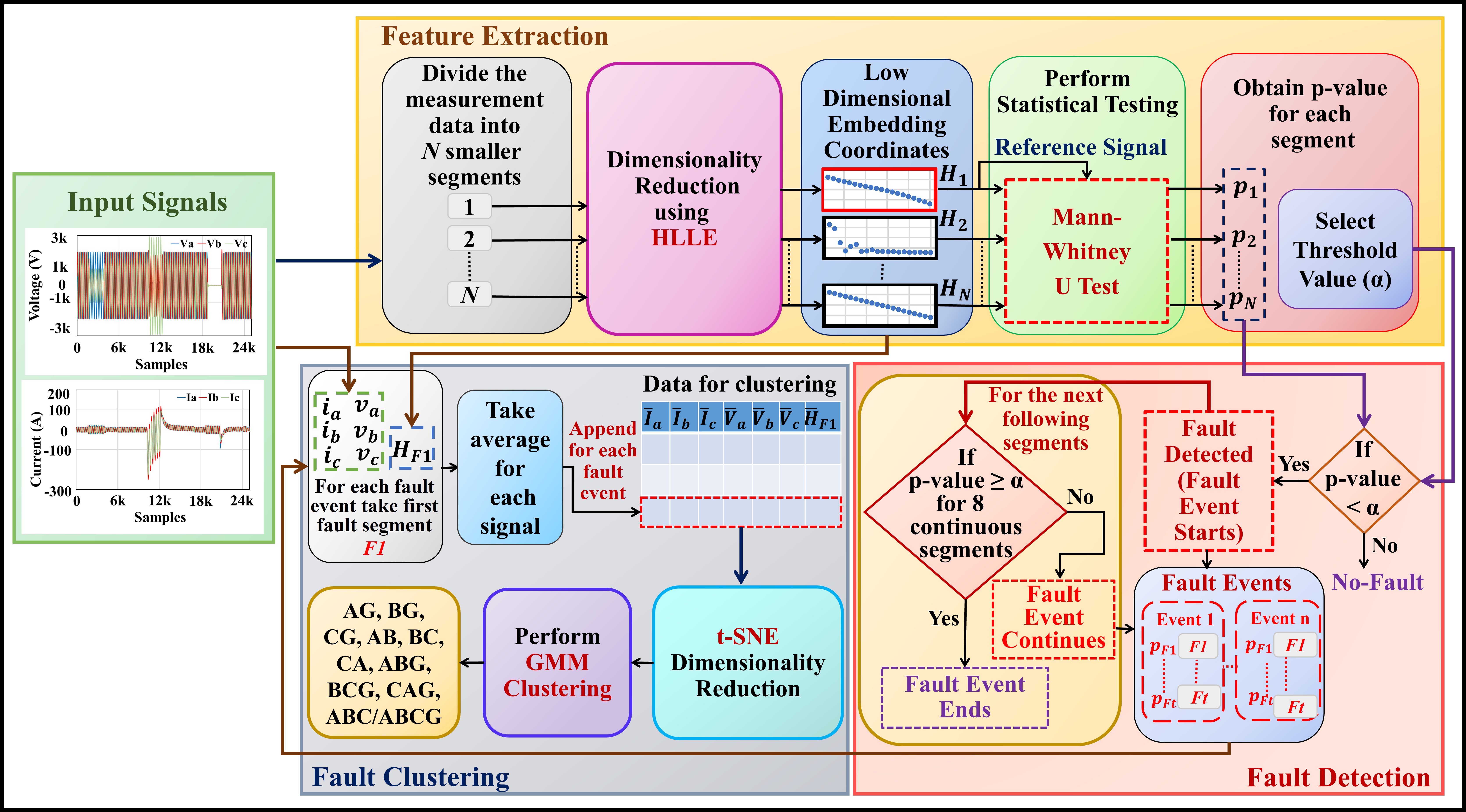}
  \caption{The complete process flow for fault detection and clustering using the proposed method; initially, the entire signal is divided into $N$ smaller segments. Then, HLLE transformation is applied to each segment, reducing the dimensionality from 6 to 1. Subsequently, the Mann-Whitney U test is used to determine the p-value for each transformed segment, using the HLLE of the first segment as the reference. These p-values are utilized to identify the starting point of the fault. It was observed that if the p-value is less than the selected threshold alpha value, a fault is detected. Once fault is detected, the following segments are considered as a fault event, and if the corresponding p-values during this event are for eight consecutive segments, then the fault event is considered to have ended. The average of all six signals consisting of three-phase voltages and currents, along with the average of HLLE values for the first fault segment, is extracted as a feature for performing clustering. After the data for clustering is extracted, it undergoes t-SNE dimensionality reduction and GMM clustering to distinctly identify the fault type.}
  \label{fig:process}
\end{figure*}

\section{Methodology}
\label{sec:Method}
The detailed step-by-step process of the proposed methodology is illustrated in Fig. \ref{fig:process}.
\subsection{Hessian locally linear embedding}

The measurement data is represented as, $\mathbf{X} = \{\mathbf{x}_1, \mathbf{x}_2, \ldots, \mathbf{x}_M\}$,
where $\mathbf{x}_i \in \mathbb{R}^D$ and $M$ denotes the total number of data points in the D-dimensional space. It is divided into $N$ segments for further processing. Each data point $\mathbf{x}_i$ is associated with its k-nearest neighbors, calculated through the Euclidean distance metric,
\begin{equation}
 d(\mathbf{x}_i, \mathbf{x}_j) = \sqrt{\sum\limits_{l=1}^D (x_{il} - x_{jl})^2}   
\end{equation}

A Hessian matrix, $\mathbf{H}(f(\mathbf{x}_i))$, for each neighborhood is estimated, embodying the second-order local information of the data. Mathematically, the Hessian matrix is expressed as 
\begin{equation}
  \mathbf{H}(f(\mathbf{x}_i)) = 
\begin{bmatrix}
\dfrac{\partial^2 f}{\partial x_{i1}^2} & \cdots & \dfrac{\partial^2 f}{\partial x_{i1} \partial x_{iD}} \\
\vdots & \ddots & \vdots \\
\dfrac{\partial^2 f}{\partial x_{iD} \partial x_{i1}} & \cdots & \dfrac{\partial^2 f}{\partial x_{iD}^2}
\end{bmatrix}  
\end{equation}

The Hessian's eigenvalues and eigenvectors are then computed through the characteristic equation $\det(\mathbf{H} - \lambda \mathbf{I}) = 0$, yielding eigenvalues $\lambda_i$ and corresponding eigenvectors $\mathbf{v}_i$, represented as 
\begin{equation}
    \mathbf{H} \mathbf{v}_i = \lambda_i \mathbf{v}_i
\end{equation}

The zero eigenvalues and corresponding eigenvectors become crucial, characterized by $\mathbf{H} \mathbf{v}_i = \mathbf{0}, \quad \text{for } \lambda_i = 0$. With the significant eigenvectors identified, we compile them into the matrix $\mathbf{V}$ as 
\begin{equation}
    \mathbf{V} = \begin{bmatrix}
| & | &  & | \\
\mathbf{v}_1 & \mathbf{v}_2 & \cdots & \mathbf{v}_r \\
| & | &  & |
\end{bmatrix}
\end{equation}

where $r$ denotes the total count of these eigenvectors. The intrinsic dimensionality $d$ is a derivative of the original dimensionality $D$ and the count of zero eigenvalues $r$, given by 
\begin{equation}
    d = D - r
\end{equation}

\subsection{Application of Mann-Whitney U test}

The Mann-Whitney U test is applied between the reference segment \( R \) (i.e., the first segment from total \( N \) segments) and each of the other segments indexed by \( i \) where \( i \in \{ 2,3, \dots, N \} \). The test is formulated to compute the p-value, assessing the null hypothesis \( H_0 \) that states there is no significant difference between the distributions of the two compared segments. The U statistic for the test is computed as
\begin{equation}
MW_U = k_1 \cdot k_2 + \frac{k_1 \cdot (k_1+1)}{2} - R_1
\end{equation}

where \( k_1 \) is the size of segment \( R \), \( k_2 \) is the size of the compared segment from \( N \), and \( R_1 \) is the sum of ranks in segment \( R \).

The p-values \cite{p_value} derived from the Mann-Whitney U test for each time segment are presented as a time series to allow both visual and quantitative evaluations of the system's state over time. A fault is identified when there is a significant deviation of the p-value from a reference value. 

A threshold, $\alpha$, is established, often determined through empirical or statistical means, serving as a baseline for anomaly detection. Mathematically, the condition for fault detection is articulated as
\begin{equation}
p_N < \alpha
\end{equation}

where $p_N$ represents the p-value for segment $N$. Crossing this threshold signifies a potential fault in the system.
The utilization of a time series representation of p-values aids in capturing the dynamic behavior of the system, enhancing the accuracy and speed of fault identification. This approach helps maintain the continuous reliability and integrity of the power system.

\subsection{t-SNE with Gaussian Mixture Model Clustering}

For clustering the detected fault events, we use t-distributed stochastic neighbor embedding in combination with Gaussian mixture model clustering. This approach is employed to explore and analyze HD data effectively while preserving the inherent relationships among data points.

Given a set of $N$ points $X = \{n_1, n_2, \ldots, n_N\}$ in the HD space, the first step is to compute the pairwise similarities between points. We calculate the conditional probability $cp_{j|i}$ that a point $n_j$ is picked as a neighbor of point $n_i$, assuming neighbors are picked as per their probability density under a Gaussian distribution centered at $n_i$. The conditional probability $cp_{j|i}$ is given by
\begin{equation}
cp_{j|i} = \dfrac{\exp\left(-\|n_i - n_j\|^2 / 2\sigma_i^2\right)}{\sum\limits_{k \neq i} \exp\left(-\|n_i - n_k\|^2 / 2\sigma_i^2\right)}
\end{equation}

where \(\|n_i - n_j\|\) denotes the Euclidean distance between \(n_i\) and \(n_j\), and \(\sigma_i\) is the variance of the Gaussian centered at \(n_i\). We then symmetrize the conditional probabilities to obtain the joint probabilities,
\begin{equation}
cp_{ij} = \dfrac{cp_{j|i} + cp_{i|j}}{2N}
\end{equation}

In the LD space, we aim to find a set of points \(Y = \{m_1, m_2, \ldots, m_N\}\) that reflects the similarities in \(P\). We measure the similarities \(sq_{ij}\) between points in the LD space using a Student’s t-distribution with one degree of freedom (or Cauchy distribution),
\begin{equation}
sq_{ij} = \dfrac{(1 + \|m_i - m_j\|^2)^{-1}}{\sum\limits_{k \neq l} (1 + \|m_k - m_l\|^2)^{-1}}
\end{equation}

The t-SNE minimizes the Kullback-Leibler (KL) divergence between the two distributions \(P\) and \(Q\), which measures the disparity between the HD and LD probability distributions. The KL divergence is defined as
\begin{equation}
C = \sum\limits_{i \neq j} cp_{ij} \log\dfrac{cp_{ij}}{cq_{ij}}
\end{equation}

The cost function \(C\) is minimized using gradient descent. The gradients of \(C\) with respect to the points \(m_i\) are calculated and used to iteratively update the positions of the points in the LD space. Upon the convergence of the gradient descent, the points in the LD space \(Y = \{m_1, m_2, \ldots, m_N\}\) provide a visual representation that approximates the pairwise similarities of the original HD data points. Typically, these are mapped onto a 2D or 3D space for visualization.

\begin{table*}[b]
\caption{Clustering Evaluation Metrics}
\renewcommand{\arraystretch}{1.6}
\begin{tabular}{>{\arraybackslash}m{2.1cm}>{\arraybackslash}m{10.5cm}>{\centering\arraybackslash}m{5cm}}
\hline
\textbf{Metric} & \textbf{Significance} & \textbf{Formula} \\
\hline
\vspace{0.1cm}Mutual Information\vspace{0.1cm}  & \vspace{0.1cm}Measures the amount of information shared between true and predicted clusterings.\vspace{0.1cm} & \vspace{0.1cm}$MI(X, Y) = H(X) + H(Y) - H(X, Y)$ \vspace{0.1cm}\\
\vspace{0.1cm}Adjusted Mutual Information\vspace{0.1cm} & \vspace{0.1cm}Measures the mutual information adjusted for chance. Higher values indicate better clustering or partitioning.\vspace{0.1cm} & \vspace{0.1cm} $AMI(X, Y) = \hspace{3cm} $ $\dfrac{MI(X, Y) - E(MI(X, Y))}{\max(H(X), H(Y)) - E(MI(X, Y))}$ \vspace{0.1cm} \\
Rand Index & Measures the similarity between two clusters, where a high value shows better agreement between clusters. & $RI(X, Y) = \dfrac{TP + TN}{TP + FP + FN + TN}$ \\
Adjusted Rand \hspace{0.1cm}  Index & Measures the similarity between two data clusterings. The value ranges between 1 to 1, where high values show better agreement between clusters. & $ARI(X, Y) = \hspace{3cm} $ $ \dfrac{RI - \text{Expected\_RI}}{\max(RI_{\text{max}} - \text{Expected\_RI}, 0)}$ \\
Completeness Score & Assesses the degree to which all data points within a given true class are grouped together in a single cluster. & $CS(X, Y) = \dfrac{H(Y|X)}{H(Y)}$ \\
Homogeneity Score & Quantifies the degree to which each cluster exclusively comprises data points belonging to a singular class. & $HS(X, Y) = 1 - \dfrac{H(Y|X)}{H(Y)}$ \\
\hline
\end{tabular}\\

\vspace{2mm}
$H(X)$ and $H(Y)$ represent the entropies of $X$ and $Y$, respectively.\\ $H(Y|X)$ denotes the conditional entropy of $Y$ given $X$.\\
$TP$ stands for True Positives, $FP$ for False Positives, $FN$ for False Negatives, and $TN$ for True Negatives.
\label{tab:cluster_metrics}
\end{table*}

\subsection{GMM Clustering}

Following the dimensionality reduction with t-SNE, GMM clustering is applied to identify underlying clusters within the lower-dimensional data. The objective is to group these events based on their characteristics, specifically the input features such as the p-value statistics, current, and voltage magnitudes. GMM is an unsupervised clustering algorithm employed to achieve this clustering task. GMMs are known for their ability to create ellipsoidal-shaped clusters, leveraging the expectation-maximization (EM) algorithm to estimate probability densities. Each cluster within a GMM is characterized by a Gaussian distribution, defined by its covariance matrix and mean. GMMs provide an enhanced quantitative assessment of data fitness for the number of clusters, utilizing the mean and covariance. They are given as a linear combination of  Gaussian probability distribution.

The formulation of a GMM is given as

\begin{equation}
    p(X) = \sum\limits_{t=1}^T \pi_t \mathcal{N}(X | \mu_t, \Sigma_t)
\end{equation}

where, $T$ signifies the total number of components in the GMM model. The mixing coefficient, $\pi_t$ helps in the estimation of the density of each Gaussian component. Each individual component, $t$, is defined through a Gaussian distribution with a mean, $\mu_t$, a covariance matrix, $\Sigma_t$, and an associated mixing coefficient, $\pi_t$.

The GMM algorithm involves an iterative process of estimating the parameters of Gaussian distributions that best fit the data. It consists of two main steps: Expectation (E-step) and Maximization (M-step). In the E-step, the algorithm estimates the expectation of the data points belonging to each Gaussian cluster. Let \(X = \{n_1, n_2, \ldots, n_N\}\) be the set of data points, and \(K\) be the number of Gaussian clusters. Each data point \(n_i\) is assigned to a Gaussian cluster with a probability given by

\begin{equation}
w_{ik} = \dfrac{\pi_k \mathcal{N}(n_i | \mu_k, \Sigma_k)}{\sum\limits_{j=1}^K \pi_j \mathcal{N}(n_i | \mu_j, \Sigma_j)}
\end{equation}

where, $w_{ik}$ is the probability of $n_i$ belonging to the $k$-th Gaussian cluster, $\pi_k$ is the mixing coefficient for the $k$-th Gaussian cluster, $\mathcal{N}(n_i | \mu_k, \Sigma_k)$ is the Gaussian distribution having $\mu_k$ as mean and $\Sigma_k$ as covariance matrix.

The M-step updates the parameters of the Gaussian distributions using the probabilities computed in the E-step. The updated parameters are calculated as follows
\begin{align}
\mu_k &= \dfrac{\sum\limits_{i=1}^N w_{ik} n_i}{\sum\limits_{i=1}^N w_{ik}} \\
\Sigma_k &= \dfrac{\sum\limits_{i=1}^N w_{ik} (n_i - \mu_k)(n_i - \mu_k)^T}{\sum\limits_{i=1}^N w_{ik}} \\
\pi_k &= \dfrac{\sum\limits_{i=1}^N w_{ik}}{N}
\end{align}

\begin{figure*}[b]
  \centering
  \includegraphics[width=7.0in]{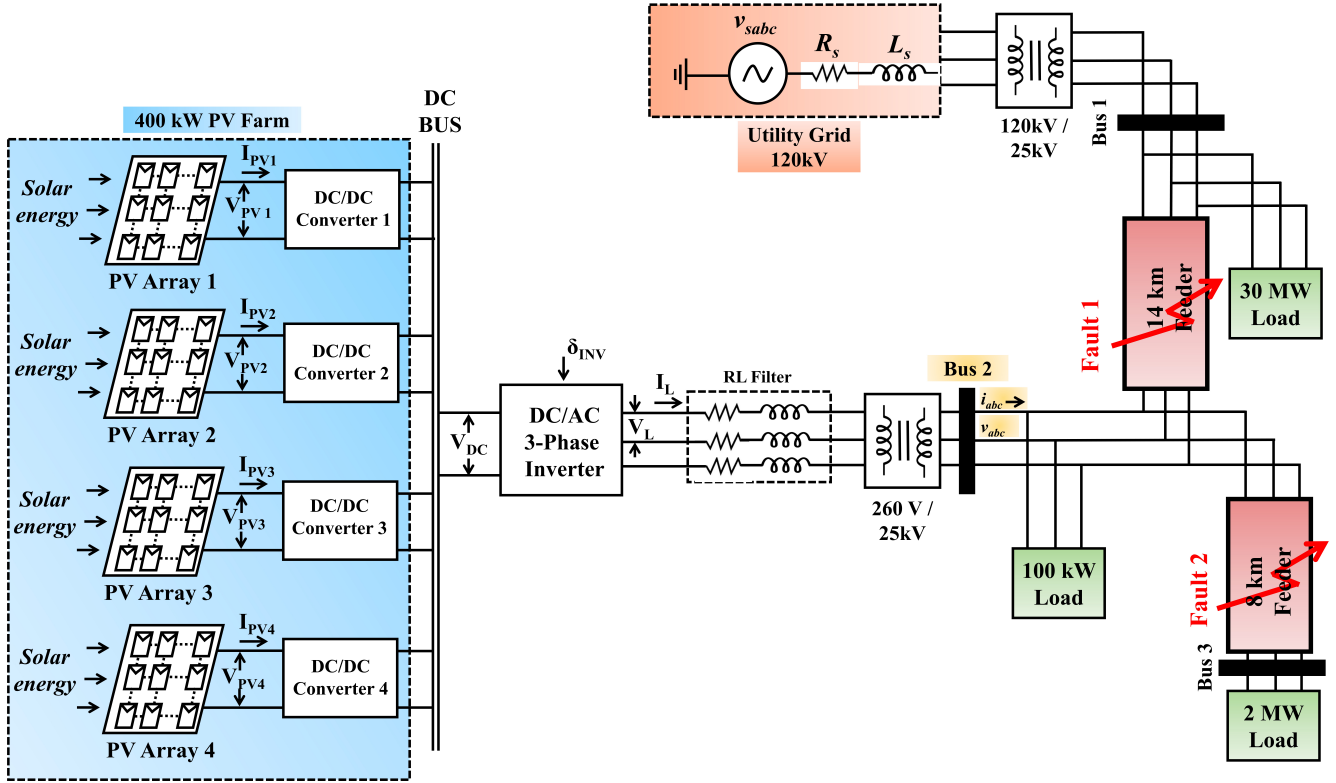}
  \caption{Schematic of the simulation model consisting of a 400 kW solar PV farm connected to the grid with three loads and two feeders.}
  \label{fig:schematic}
\end{figure*}
The updated parameters enhance the Gaussian distributions for subsequent iterations.

The E-step and M-step are repeated until the changes in the parameters or the log-likelihood fall below a threshold, indicating the convergence of the algorithm. The log-likelihood function is given by
\begin{equation}
\log p(X|\pi, \mu, \Sigma) = \sum\limits_{i=1}^N \log \left( \sum\limits_{k=1}^K \pi_k \mathcal{N}(n_i | \mu_k, \Sigma_k) \right)
\end{equation}

To evaluate the performance of the proposed clustering model, we have computed various metrics given their significance and formulas in Table \ref{tab:cluster_metrics}.

\section{Simulation and Results}
\label{sec:results}

\subsection{Simulation Model}

The schematic of the simulation model is shown in Fig. \ref{fig:schematic}. The simulation is performed in MATLAB/SIMULINK. The considered system consists of a solar PV farm of four 100 kW PV arrays connected to a distribution system. The distribution system consists of 2 feeders of 8 km and 14 km supplying 3 loads. There are 3 buses in the system and the PV farm is connected to Bus 2 where the three-phase voltage and current measurements are measured and given as input to the proposed model for fault detection and clustering.

Faults have been simulated on the two feeders for every 1 km distance; therefore, there are a total of 23 fault locations. Also, 5 different fault resistance values (0.01, 0.20, 2.00, 6.00, 10.00 ohms) are considered. Thus, there are 115 cases for each fault type. Ten various types of fault are considered; single line-to-ground fault, i.e., AG, BG, CG, double line-to-ground fault, i.e., ABG, BCG, CAG, double line fault, i.e., AB, BC, CA, and triple line-to-ground fault i.e., ABCG. The sampling rate of the simulation is 5e\textsuperscript{-5} seconds i.e., 320 samples per cycle of 60Hz frequency. Each fault is simulated for 100 milliseconds i.e., 2000 samples. Using the three-phase voltage and current measurements at Bus 2, we perform the proposed methods for fault detection and clustering.

\subsection{Results and Discussion} 
\subsubsection{Fault Detection Results}

\begin{figure*}[t]
\centering
\captionsetup[subfigure]{font=small}
\subfloat[Phase voltages.]{\includegraphics[width=0.24\textwidth, height=0.158\textwidth]{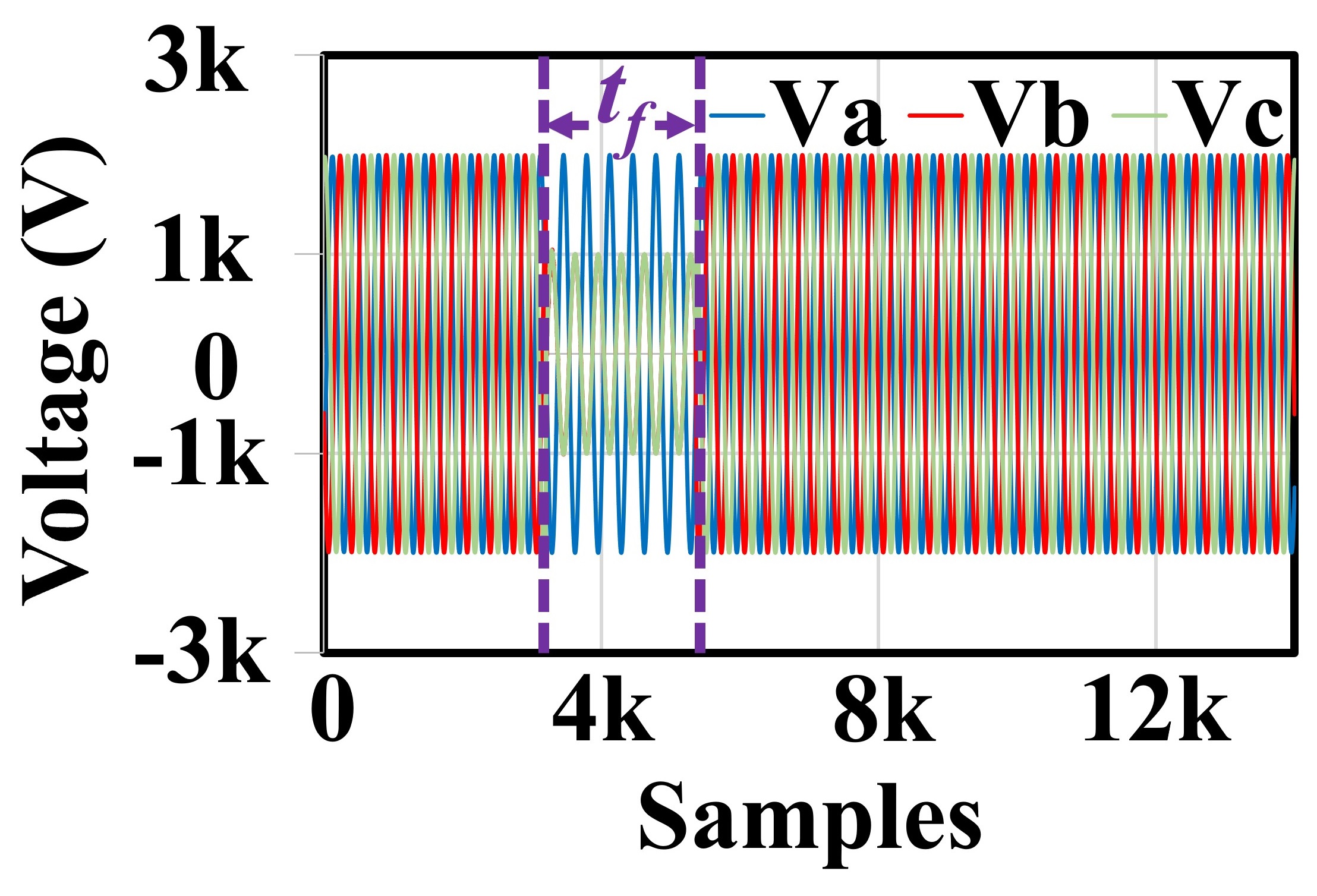}\label{fig:one_case_voltages}}\;
\subfloat[Phase currents.]{\includegraphics[width=0.24\textwidth, height=0.155\textwidth]{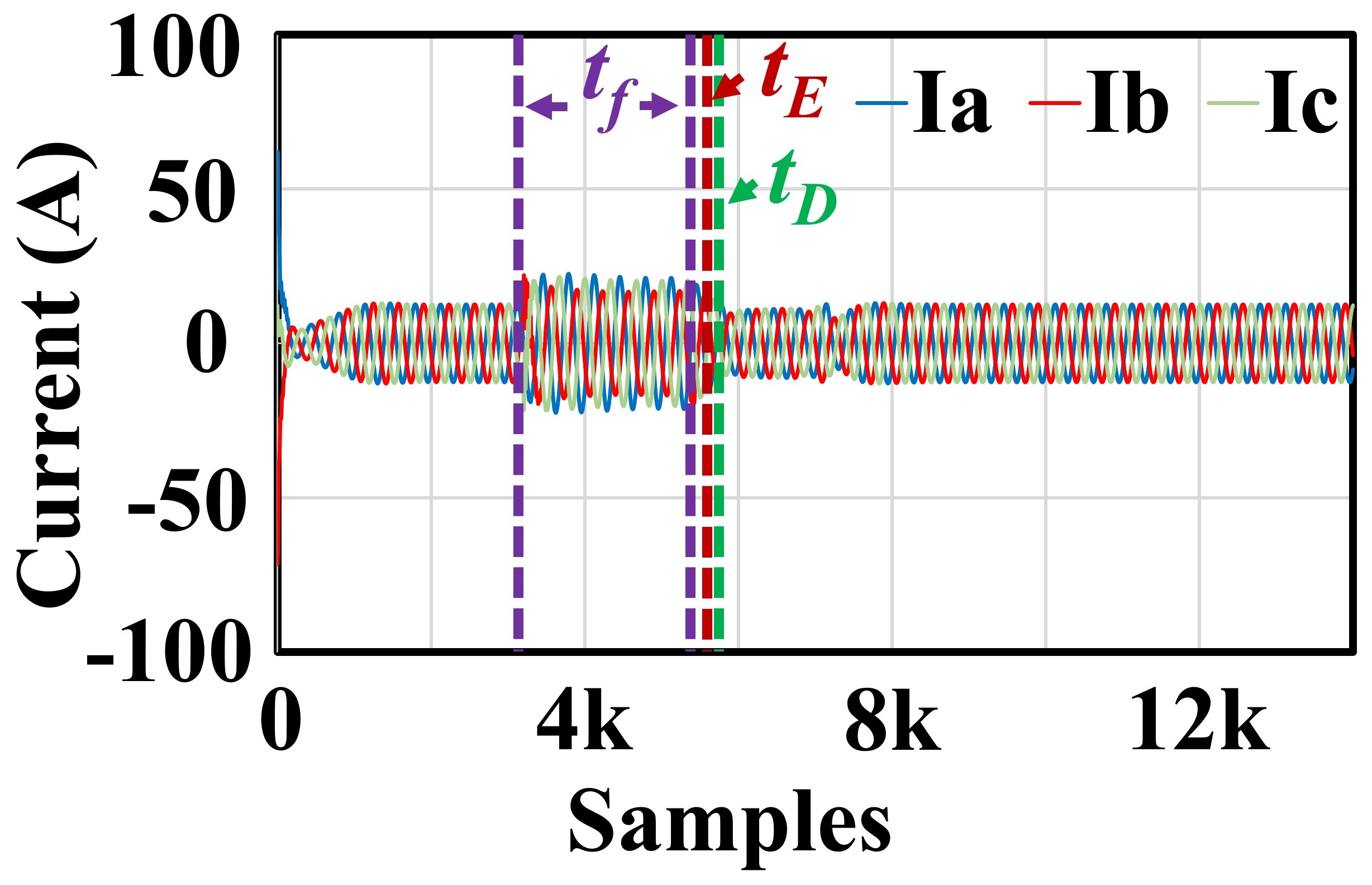}\label{fig:one_case_currents}}\;
\subfloat[p-value.]{\includegraphics[width=0.24\textwidth, height=0.155\textwidth]{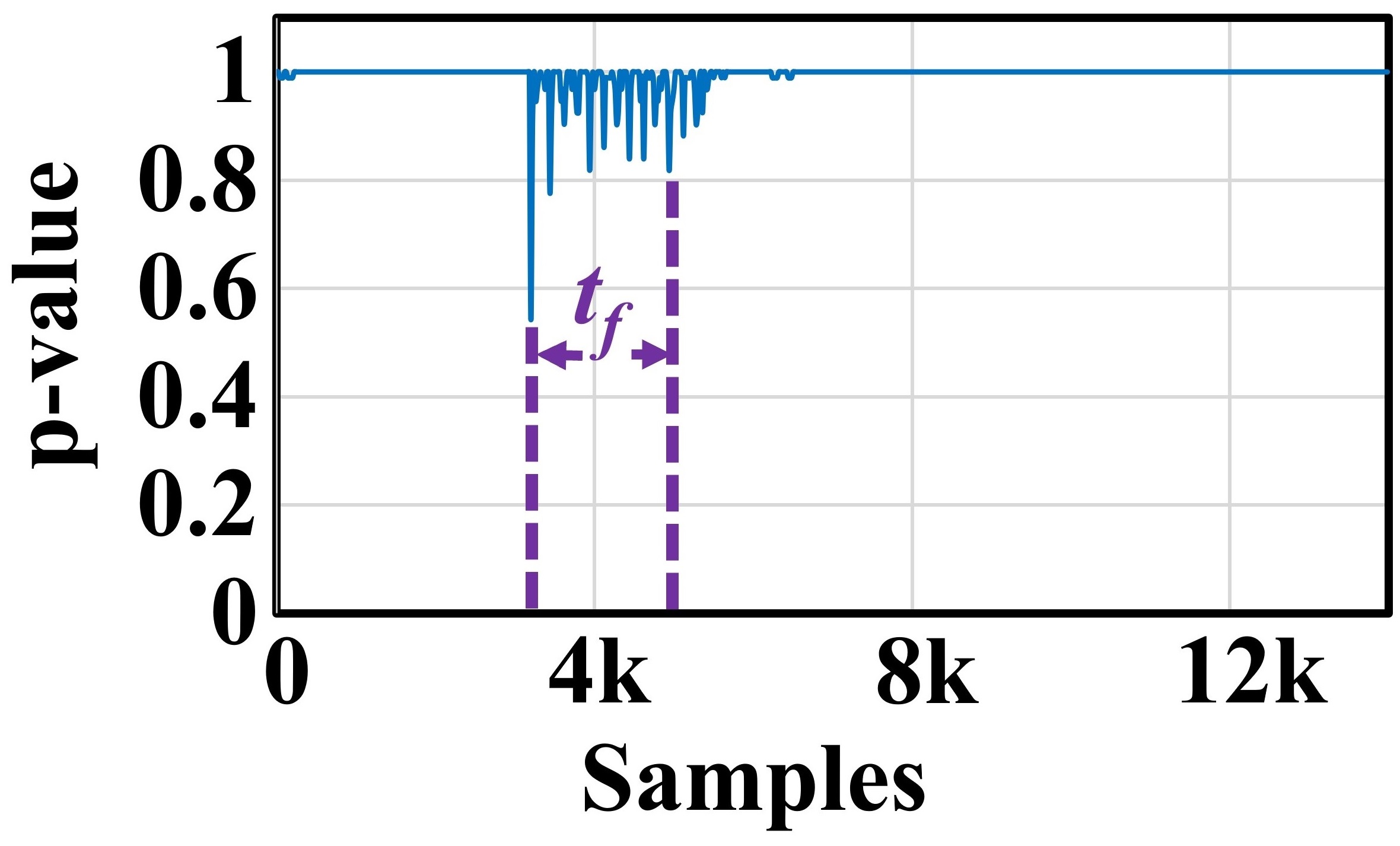}\label{fig:one_case_p_value}}\;
\subfloat[p-value (zoomed).]{\includegraphics[width=0.24\textwidth, height=0.155\textwidth]{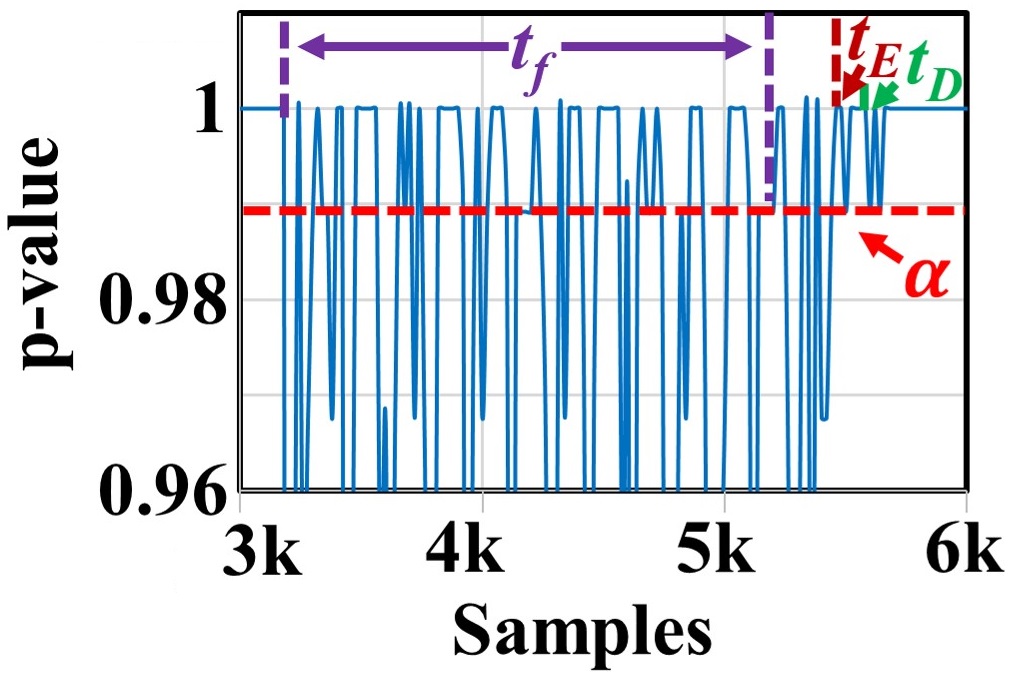}\label{fig:one_case_p_value_zoom}}
\hfill
\subfloat[1\textsuperscript{st} segment - No fault.]{\includegraphics[width=0.24\textwidth]{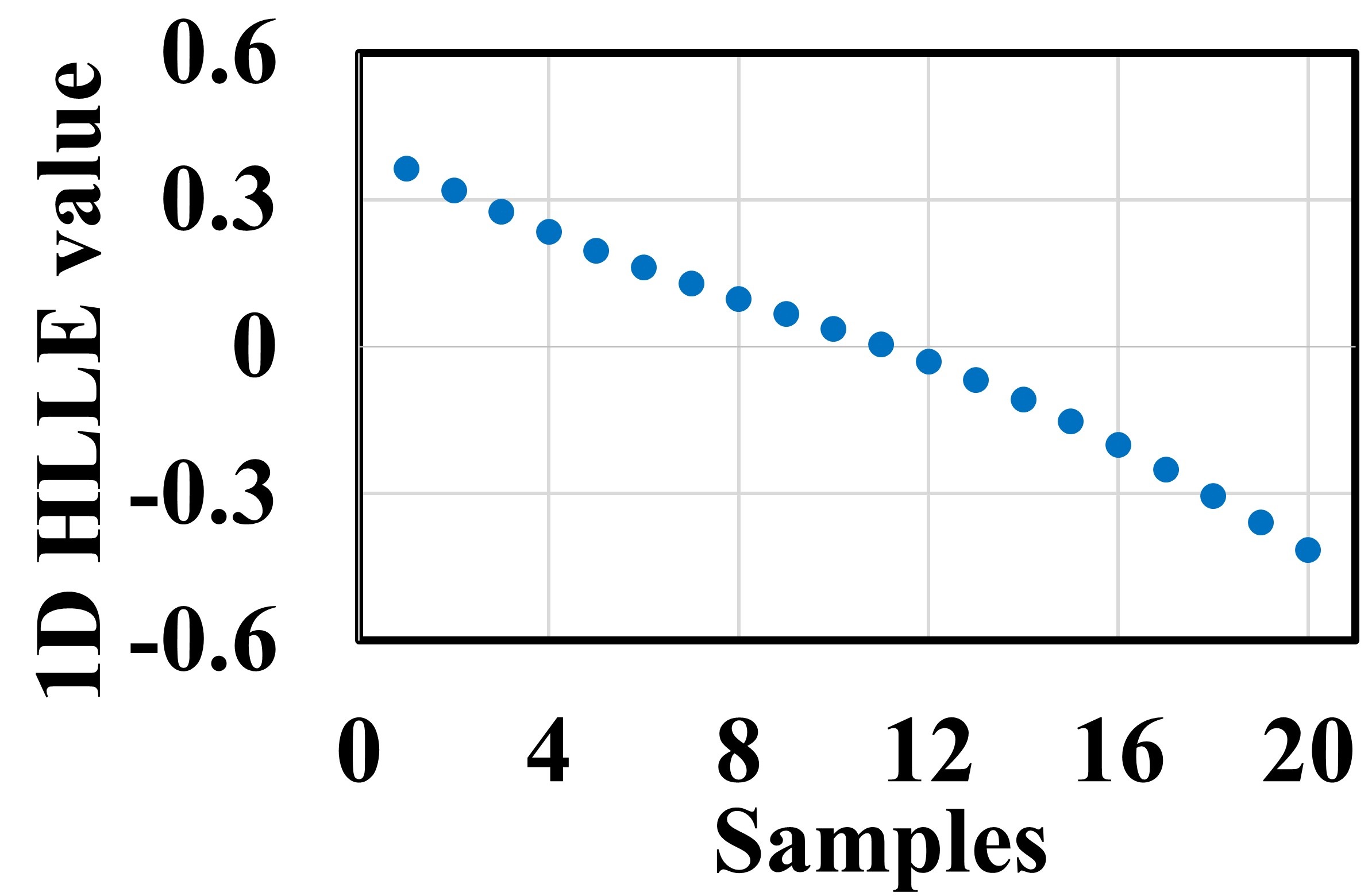}\label{fig:one_case_1_seg}}\;
\subfloat[160\textsuperscript{nd} segment - At fault instant.]{\includegraphics[width=0.24\textwidth]{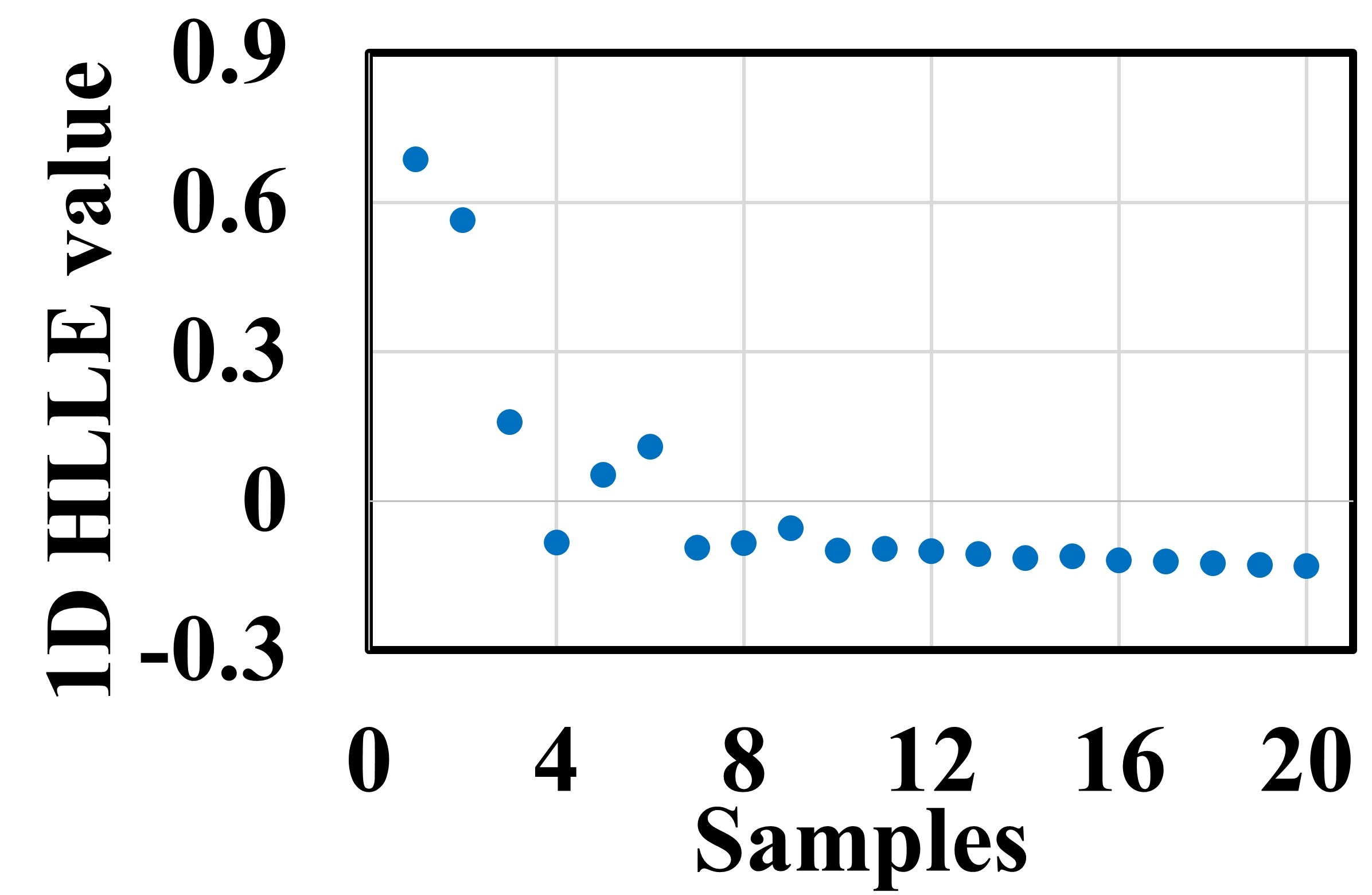}\label{fig:one_case_32_seg}}\;
\subfloat[250\textsuperscript{th} segment - During fault.]{\includegraphics[width=0.24\textwidth]{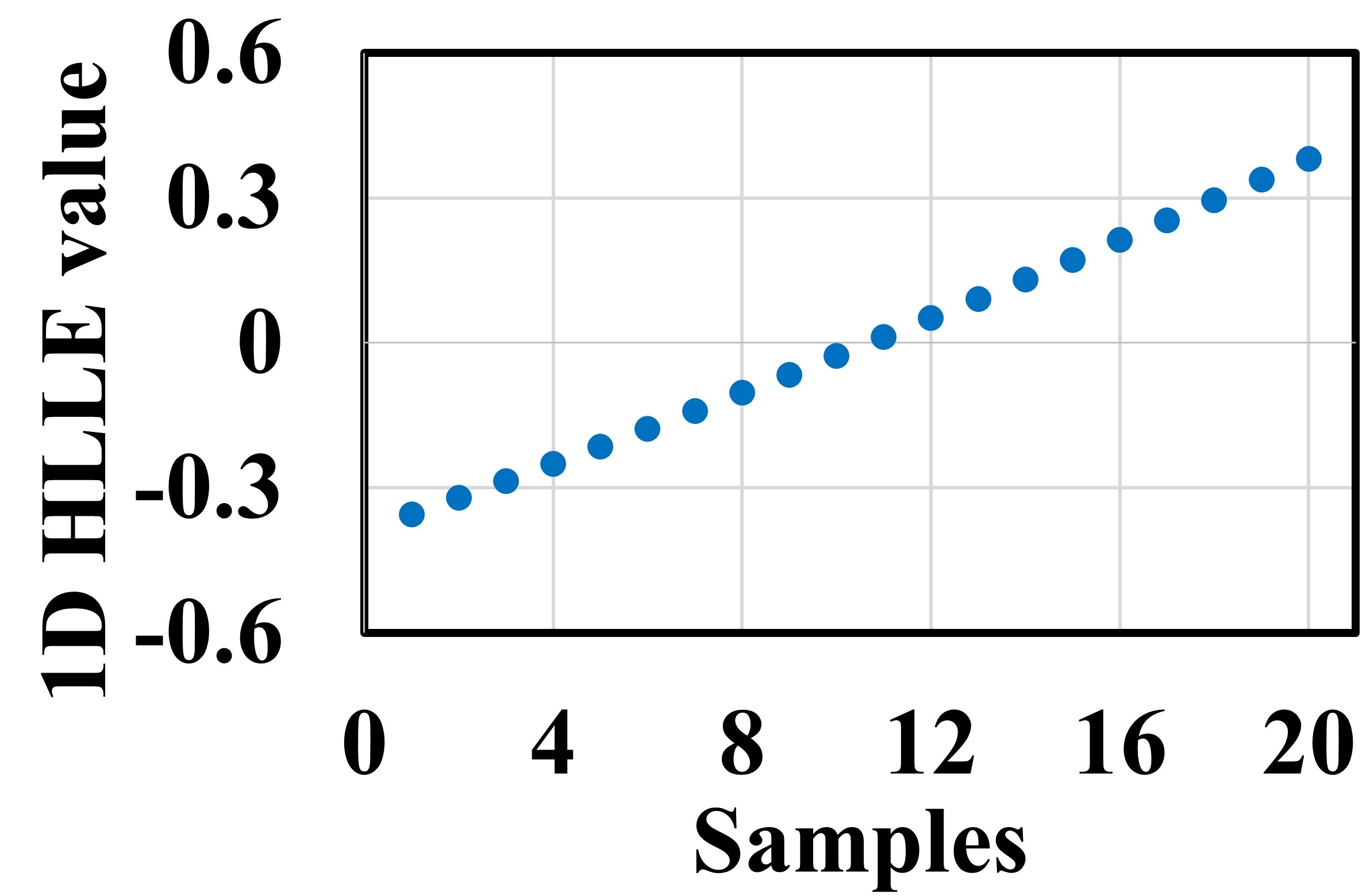}\label{fig:one_case_36_seg}}\;
\subfloat[400\textsuperscript{th} segment - After fault.]{\includegraphics[width=0.24\textwidth]{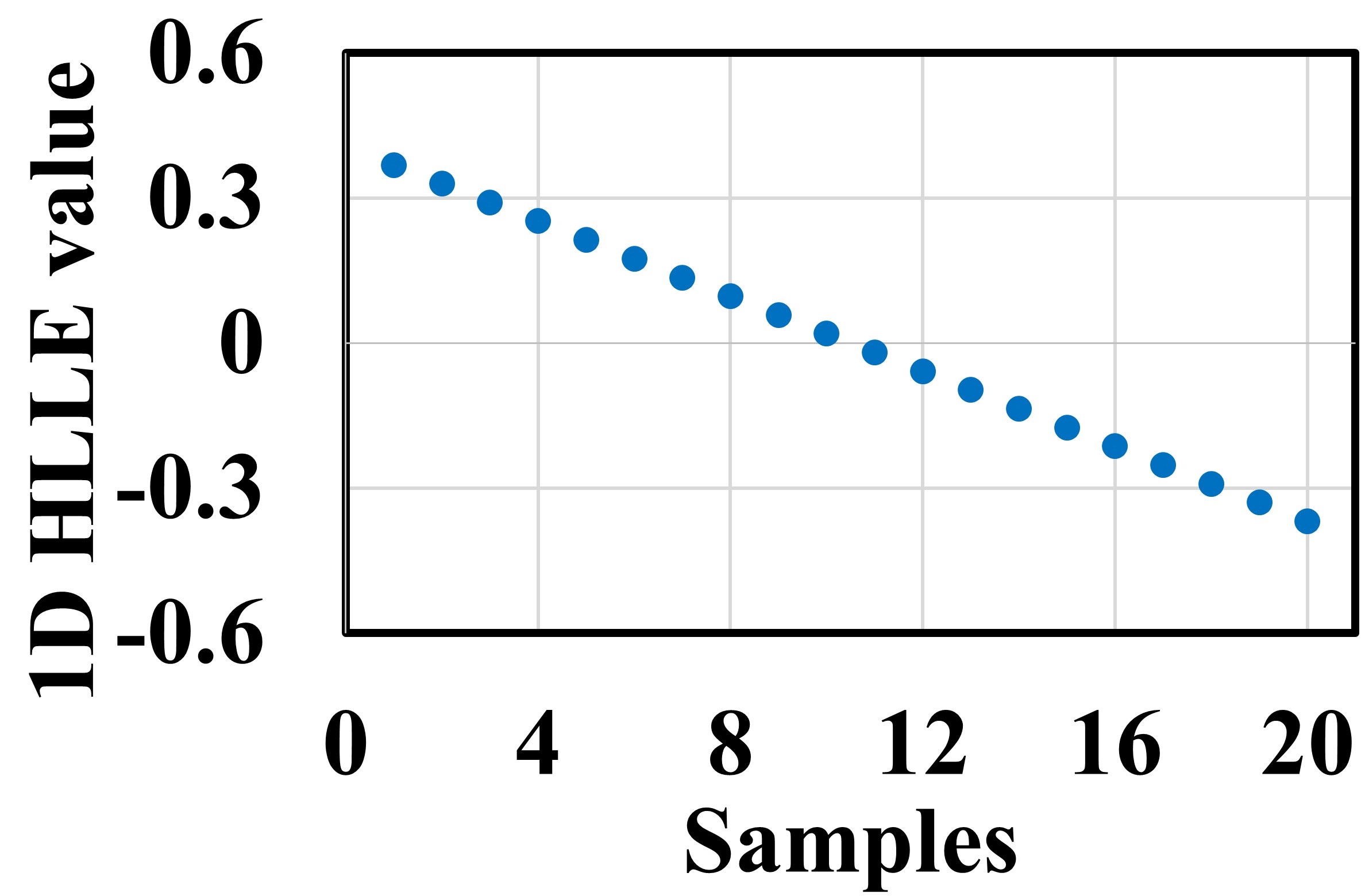}\label{fig:one_case_60_seg}}
\caption{The measurement data consists of HD data with 14000 samples for 6 columns consisting of three-phase (a) voltages and (b) currents. For each 20 samples, we perform HLLE to transform the data into a lower one-dimension. Then, we compare the transformed 1D HLLE values of 20 samples with the 1D HLLE values of reference 20 samples. The (c) p-value provides the statistical information from which fault could be detected and key inferences are observed from the (d) zoomed p-values. Every set of 20 samples is considered as segments and converted into 1D HLLE values. The reference is the (e) 1\textsuperscript{st} segment consisting of samples 1 to 20. This is compared with each segment of the data; there are a total of 700 segments for one fault case. At (f) 160\textsuperscript{th} segment, the fault has started, which results in different 1D HLLE values compared to the reference. The variations in the 1D HLLE values could be observed all throughout the fault event as seen in the (g) 250\textsuperscript{th} segment as well. Once the fault is cleared, the transformed values also return similar values of the no-fault segment as observed in (h) 400\textsuperscript{th} segment.}
\label{fig:one_case}
\end{figure*}
\begin{figure*}[h!]
\centering
\captionsetup[subfigure]{font=small}
\subfloat[Phase Voltages - AG fault.]{\includegraphics[width=0.24\textwidth, height=0.157\textwidth]{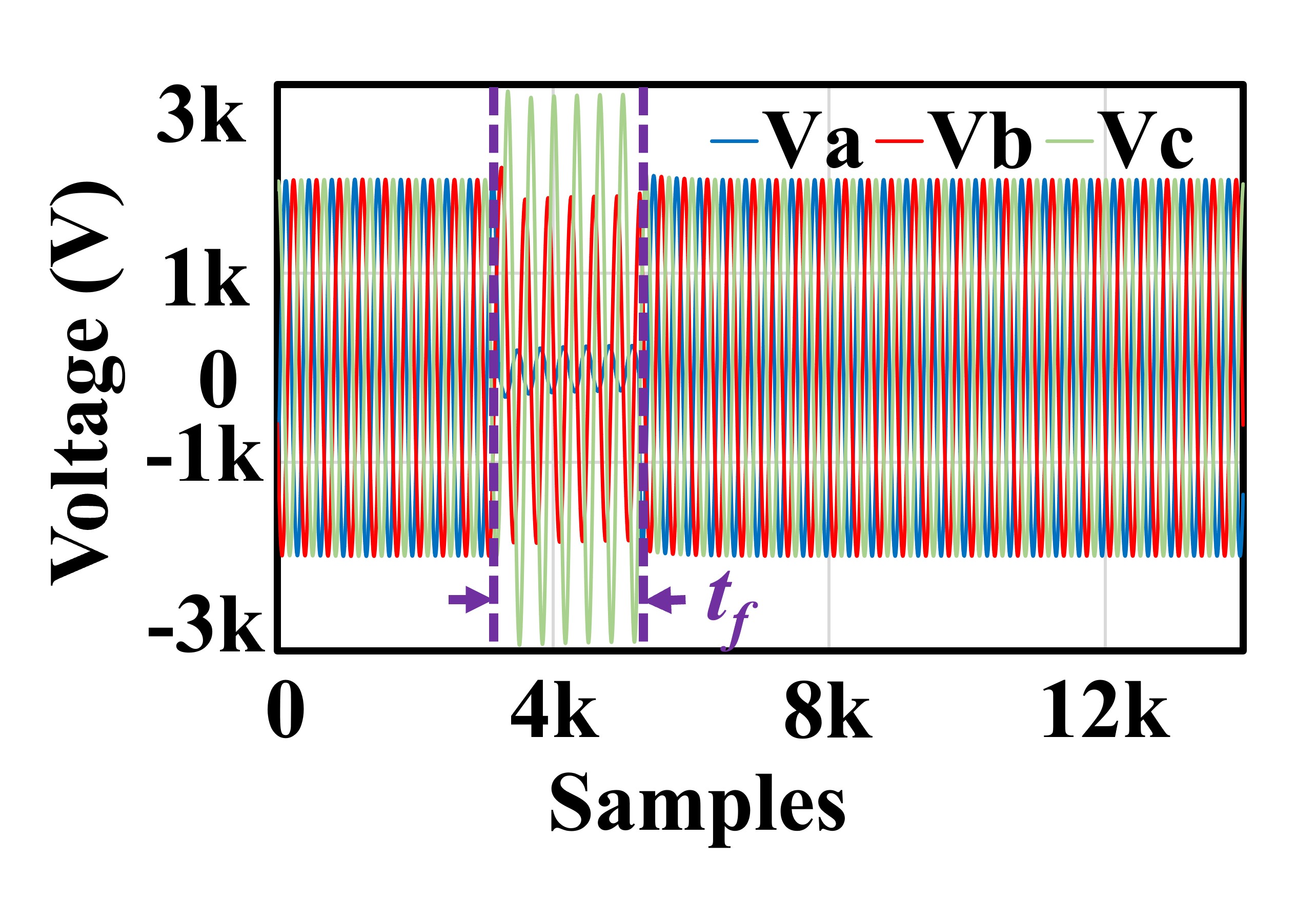}\label{fig:sub5}}\;
\subfloat[Phase Currents - AG fault.]{\includegraphics[width=0.24\textwidth, height=0.145\textwidth]{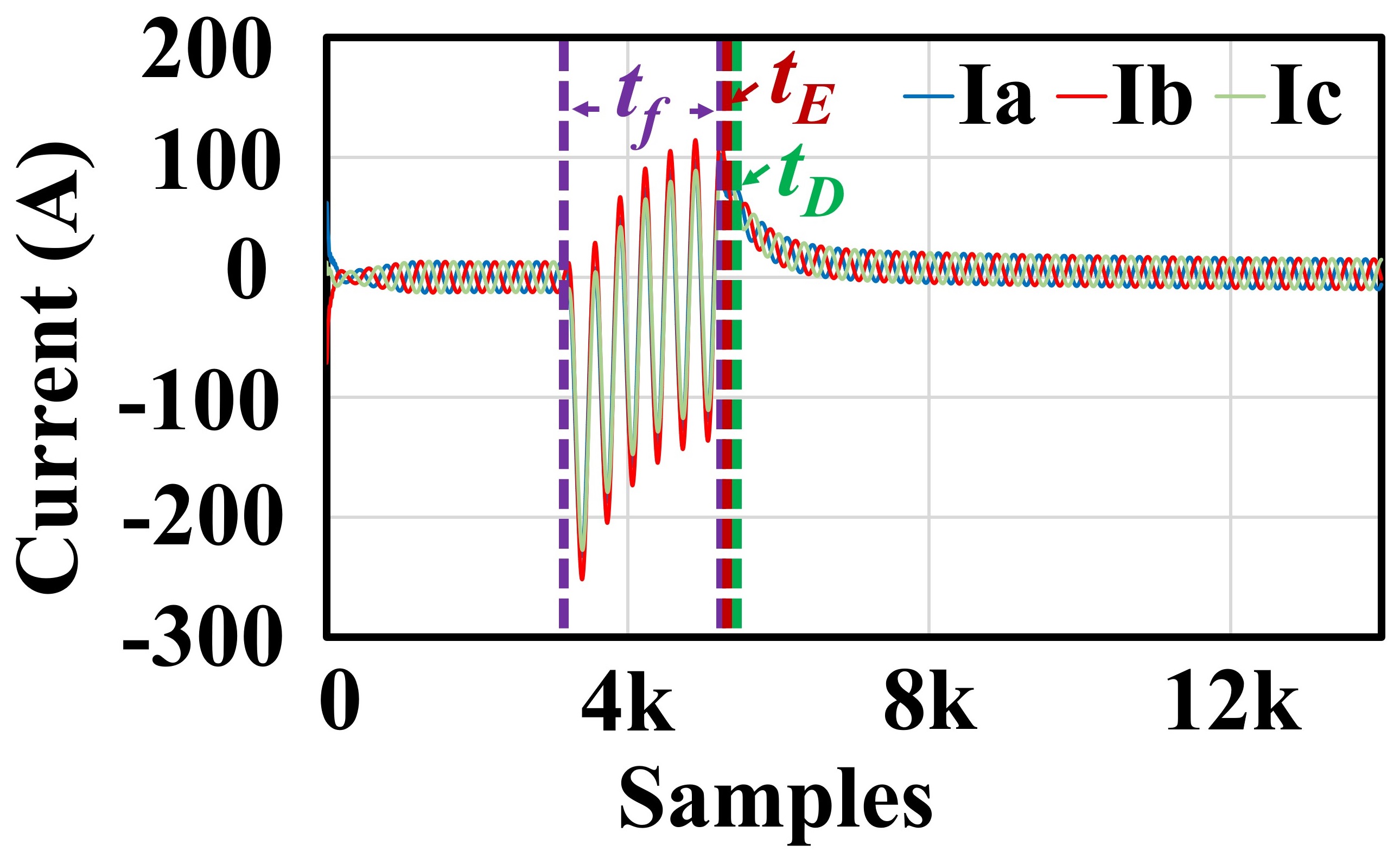}\label{fig:sub5}}\;
\subfloat[p-values - AG fault.]
{\includegraphics[width=0.24\textwidth, height=0.145\textwidth]{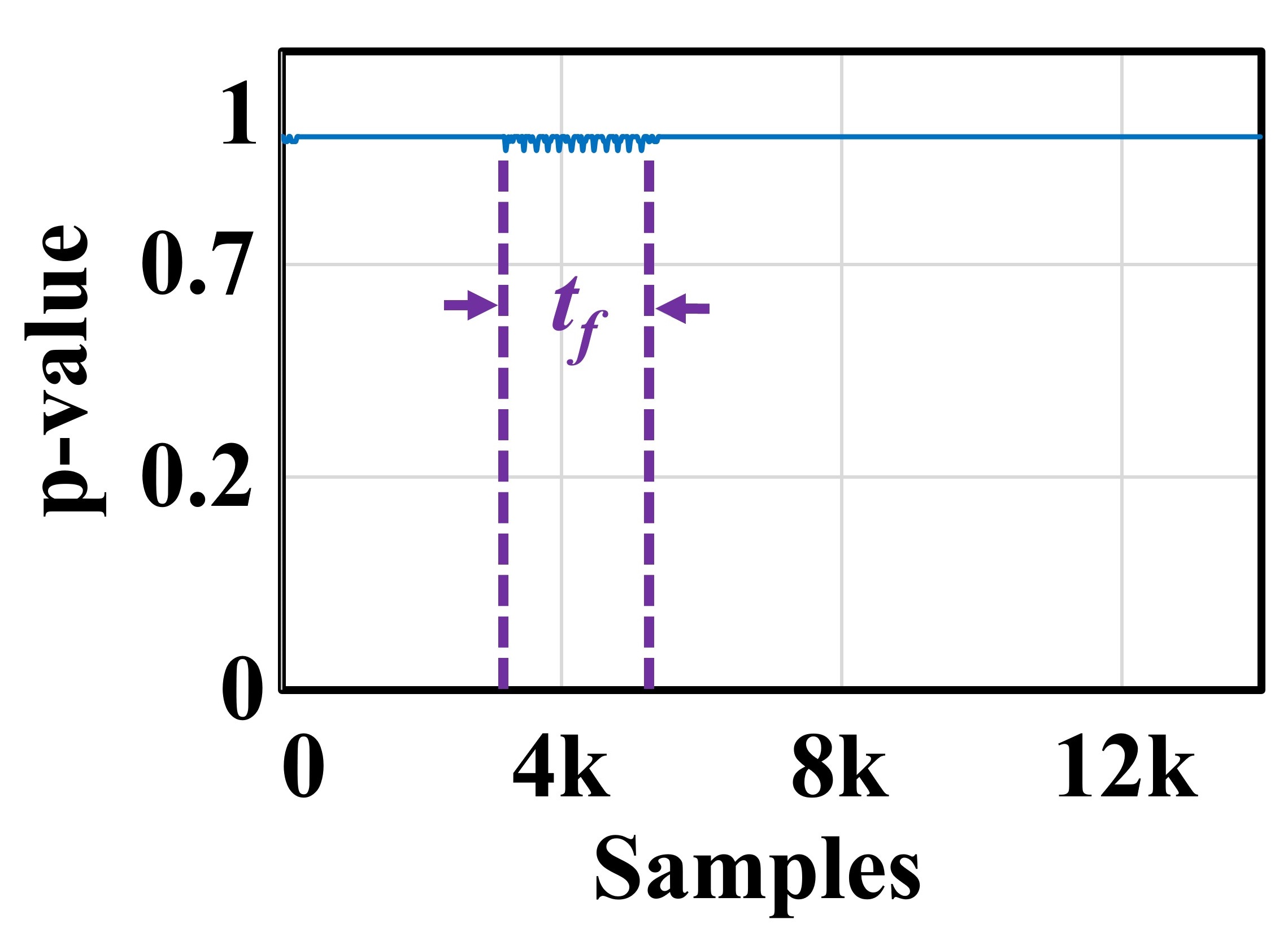}\label{fig:sub6}}\;
\subfloat[p-values (zoomed) - AG fault.]
{\includegraphics[width=0.24\textwidth, height=0.15\textwidth]{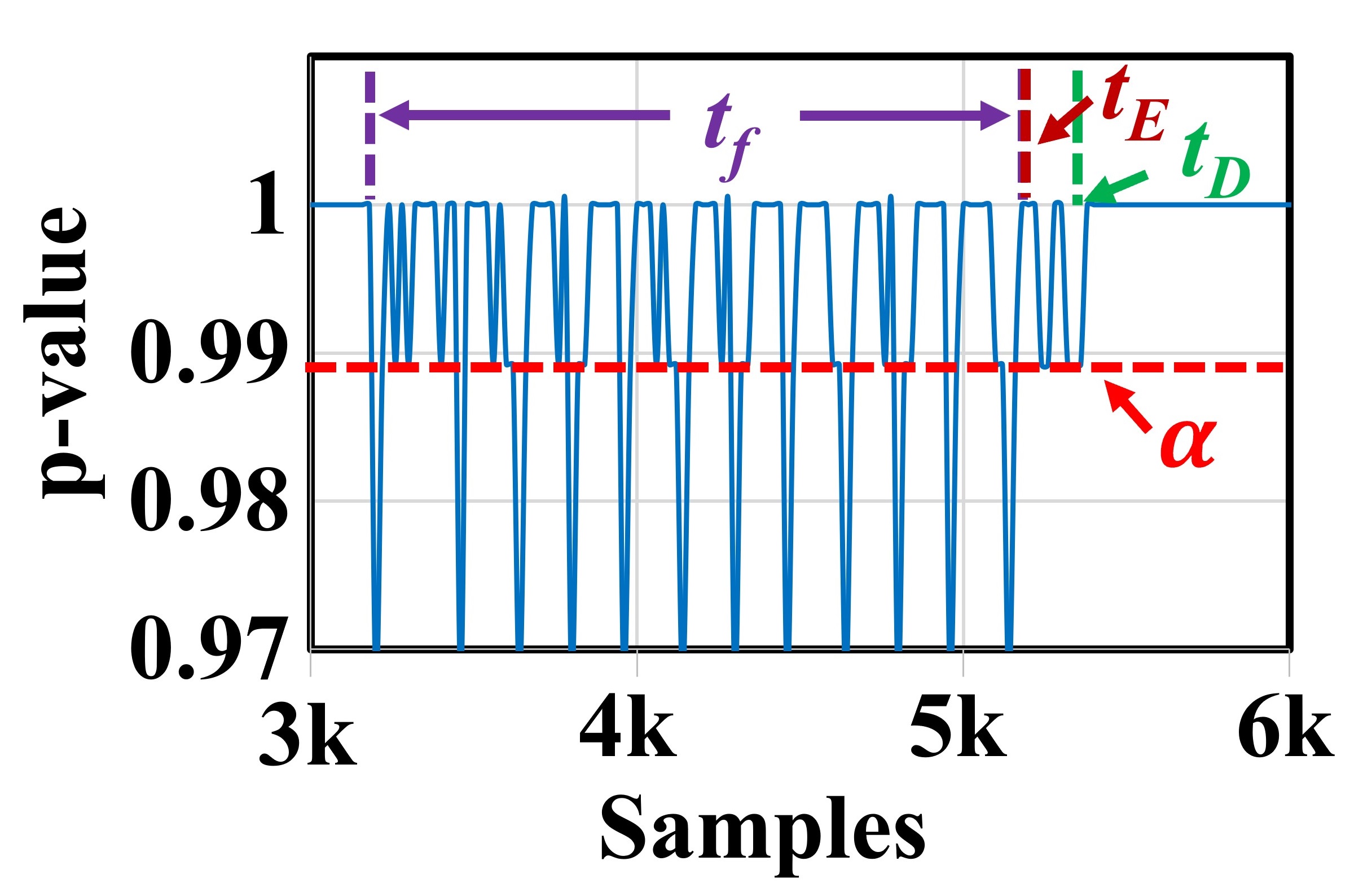}\label{fig:sub6}}
\hfill
\subfloat[Phase Voltages - ABG fault.]{\includegraphics[width=0.22\textwidth, height=0.15\textwidth]{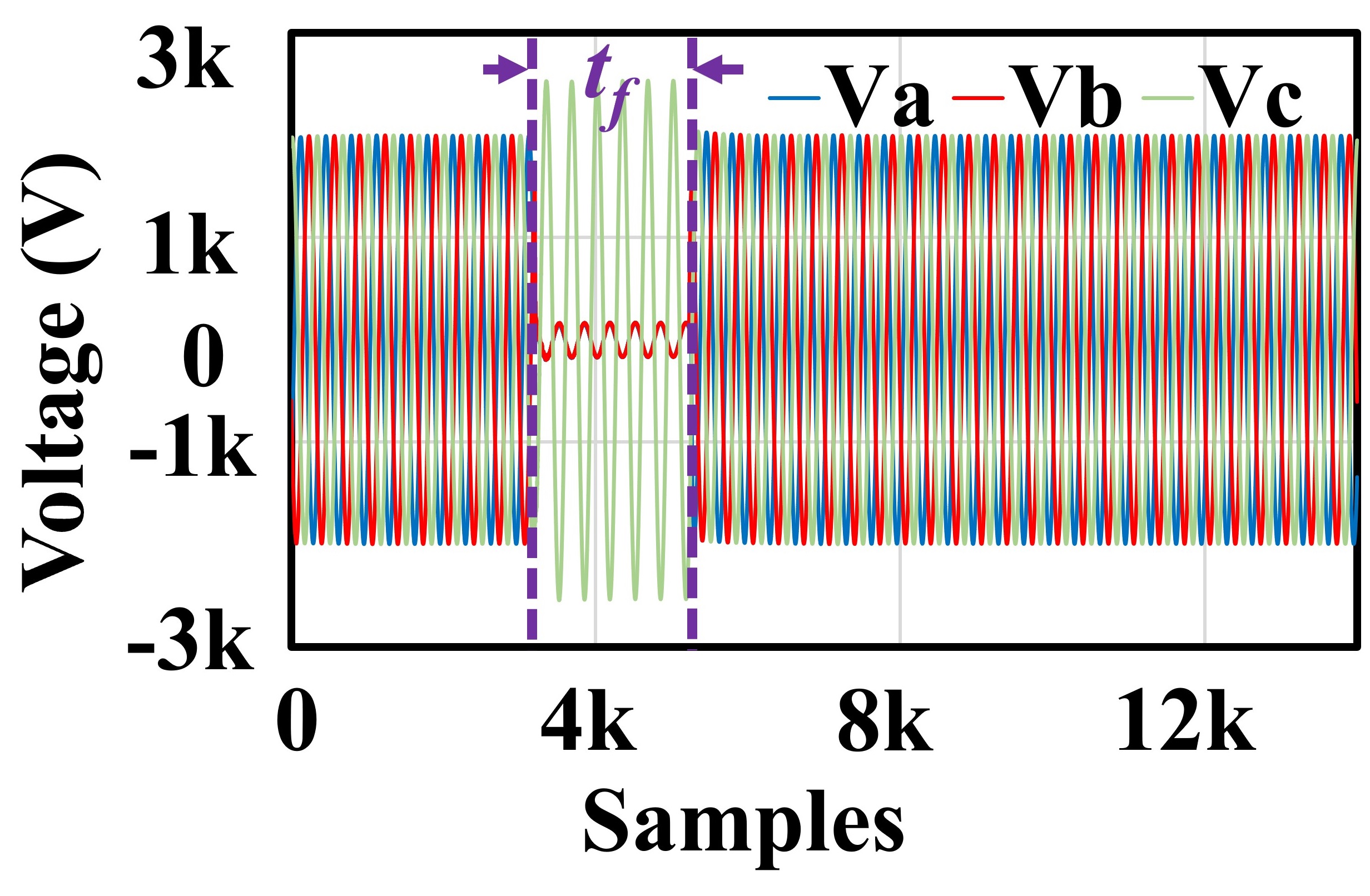}\label{fig:sub6}}\;
\subfloat[Phase Currents  - ABG fault.]{\includegraphics[width=0.24\textwidth, height=0.152\textwidth]{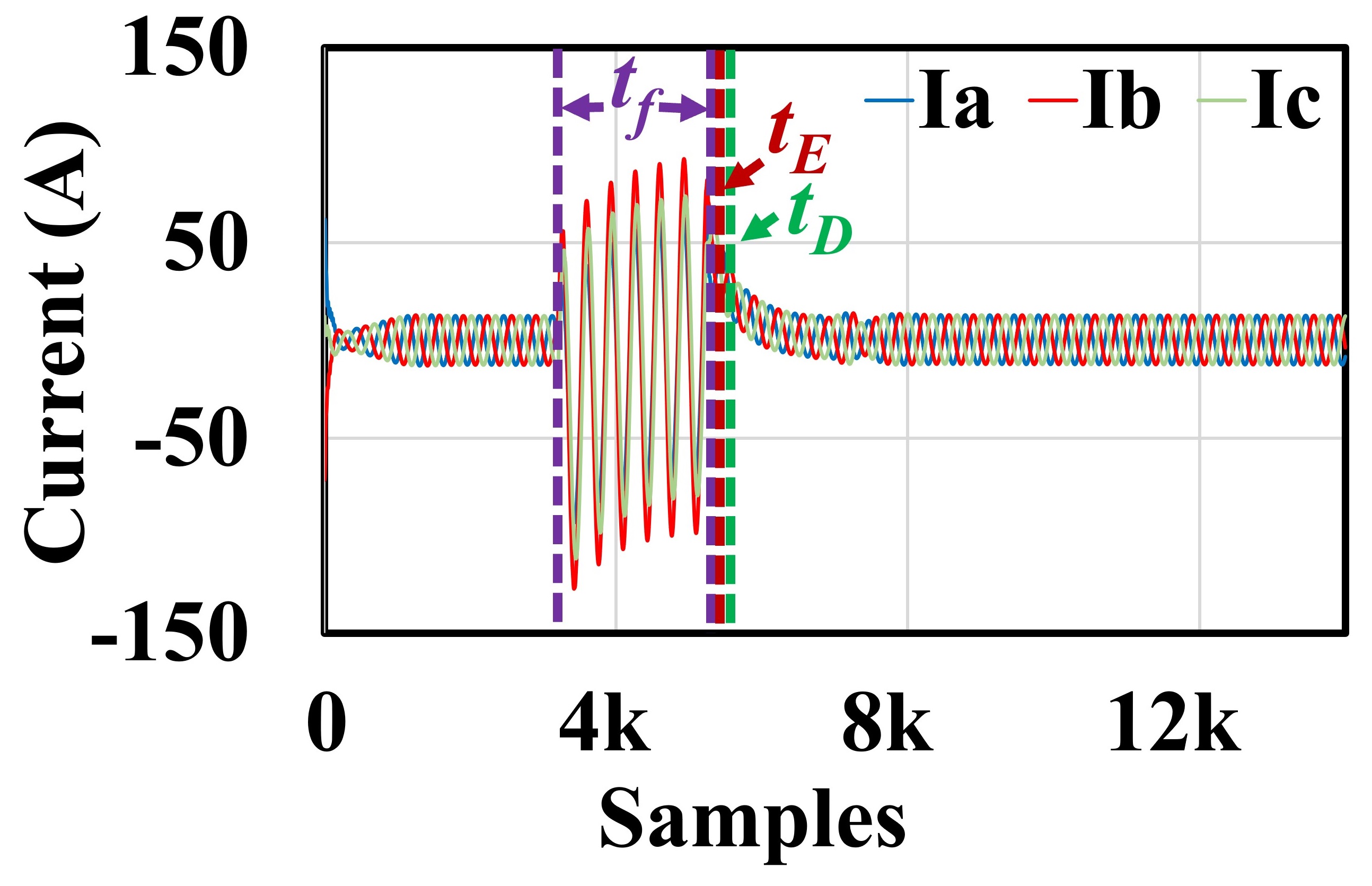}\label{fig:sub6}}\;
\subfloat[p-values - ABG fault.]
{\includegraphics[width=0.24\textwidth, height=0.15\textwidth]{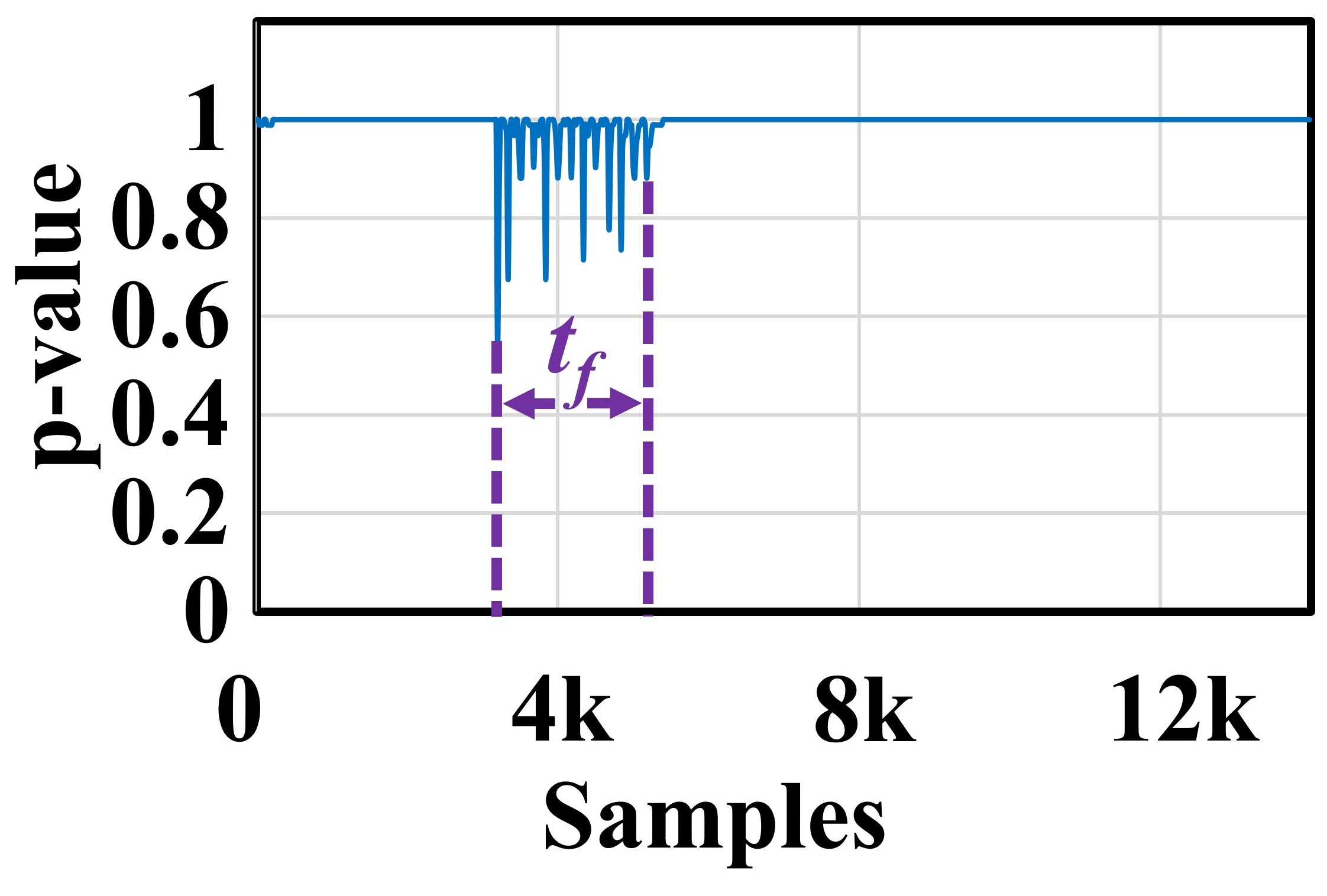}\label{fig:sub6}}\;
\subfloat[p-values (zoomed) - ABG fault.]
{\includegraphics[width=0.24\textwidth, height=0.145\textwidth]{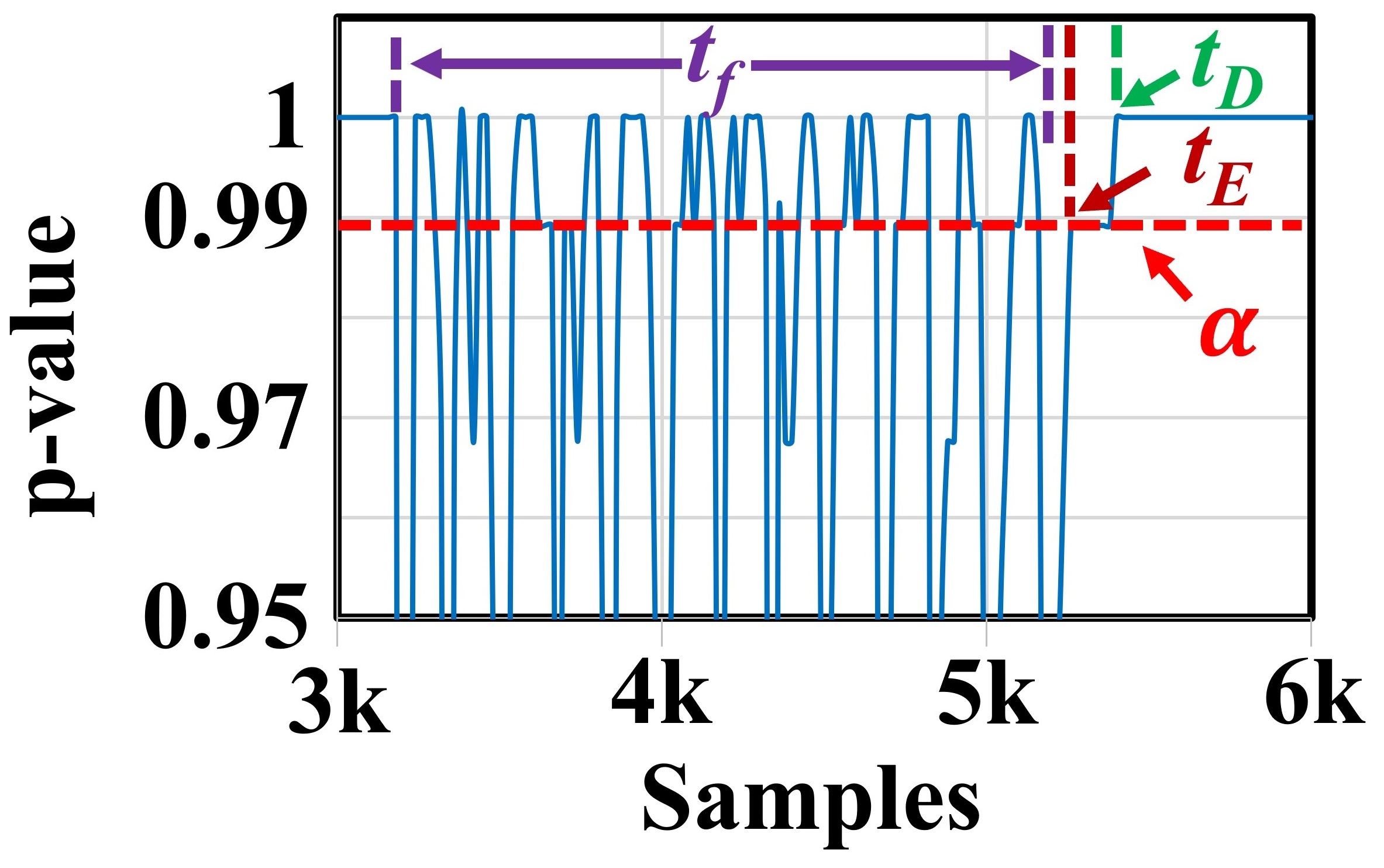}\label{fig:sub6}}
\hfill
\subfloat[Phase Voltages - TLG fault.]{\includegraphics[width=0.24\textwidth, height=0.157\textwidth]{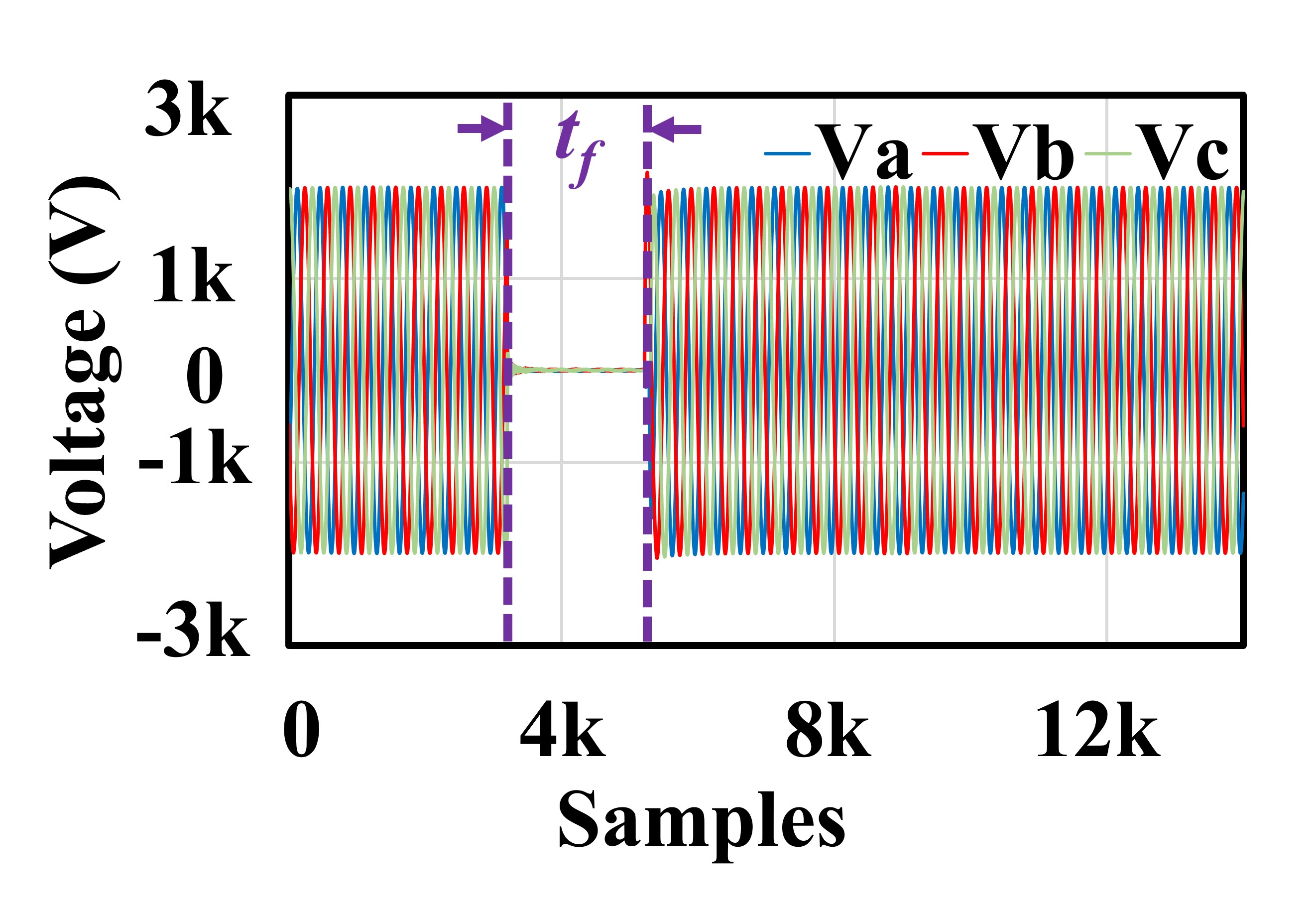}\label{fig:sub7}}\;
\subfloat[Phase Currents  - TLG fault.] 
{\includegraphics[width=0.24\textwidth, height=0.157\textwidth]{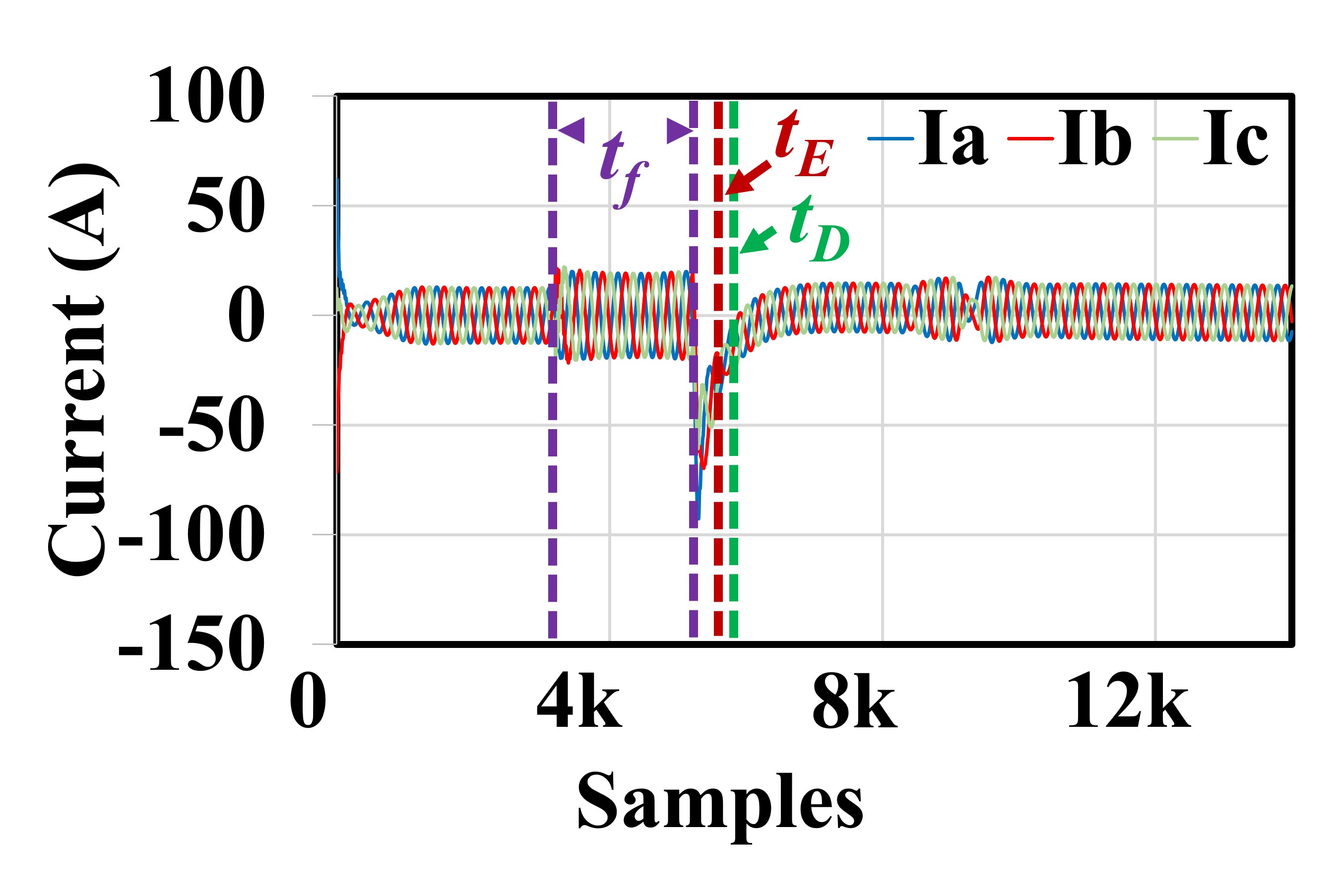}\label{fig:sub7}}\;
\subfloat[p-values - TLG fault.]
{\includegraphics[width=0.24\textwidth, height=0.15\textwidth]{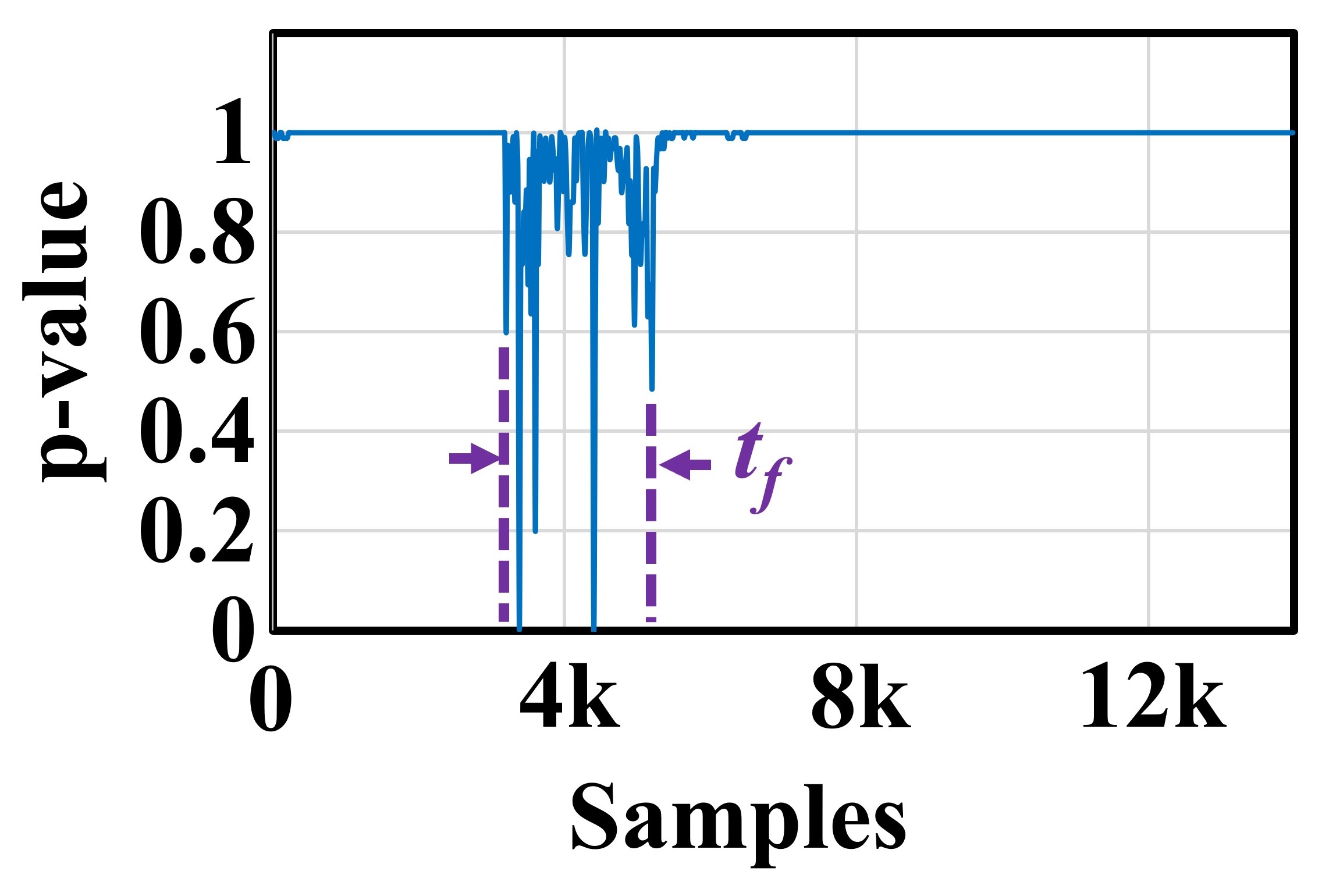}\label{fig:sub6}}\;
\subfloat[p-values (zoomed) - TLG fault.]
{\includegraphics[width=0.24\textwidth, height=0.15\textwidth]{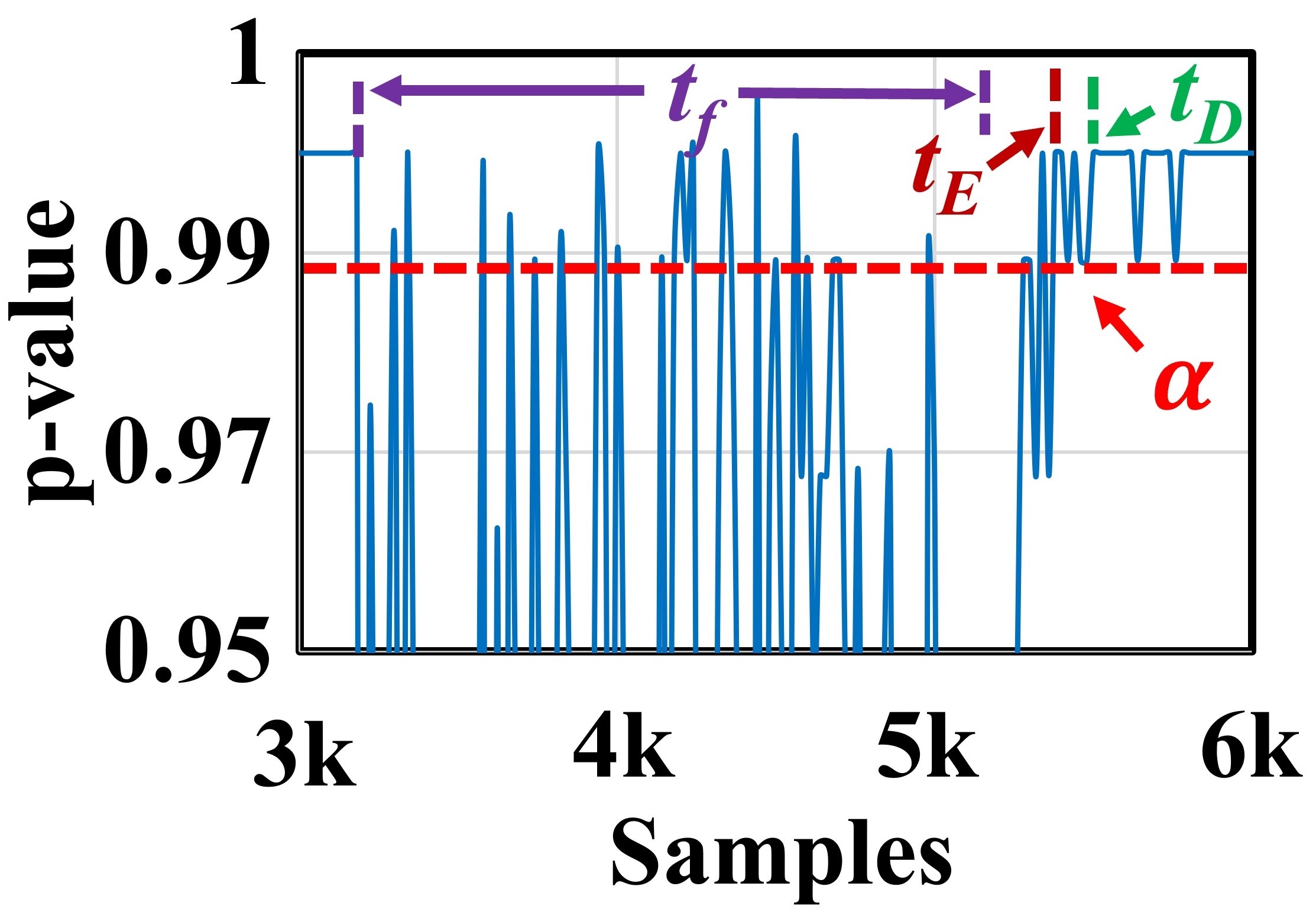}\label{fig:sub6}}

\caption{For fault cases - AG [(a), (b), (c), (d)], ABG [(e), (f), (g), (h)], TLG [(i), (j), (k), (l)], the measurements of three-phase voltages and currents are taken as input to the HLLE model where lower dimensional reduction is performed and statistical tests of each segment with the reference segment gives us p-values from which faults are detected and observations from the zoomed p-values provide the difference between in $t_f, t_E$ and $t_D$.}
\label{fig:other_cases}
\end{figure*}

The three-phase voltage and current measurements are taken as input for the detection of faults in the system. From the simulation, we have a total of 115 fault cases for each fault type, and we have concatenated all the faults into one single file to test the proposed fault detection model. Each fault case consists of 14000 samples; therefore, the concatenated single file consists of $115\times10\times14000$ samples. The first 20 samples of the concatenated file are taken as reference samples for comparing the entire data for faults. The detailed methodology is explained in Section II, and the completed process is also shown in Fig. \ref{fig:process}. We have shown the process for a BC fault case in Fig. \ref{fig:one_case}. The entire fault is simulated from 3200 samples to 5200 samples and this is considered as the fault duration $t_f$. The fault start is accurately detected from the first 20 samples (3200 to 3220) through a drop in p-value below the threshold. As there are 320 samples per cycle, we are able to detect the fault for one-sixteenth of the cycle. The p-values are sometimes above the threshold during the fault duration and even after fault clearance $t_f$ due to variations in the voltages and currents after clearance which makes it difficult to accurately identify the fault duration. Thus, the fault event is extended beyond the fault duration in some cases. For accurate fault detection, we consider the segments after the first fault detected segment; if the p-value is above the threshold for continuous eight segments, then it is considered to be the end of the fault event $t_D$. This is generally observed after a few segments after the fault is cleared at $t_E$ due to variations in voltages and currents after clearing the fault.

\begin{figure*}[t]
  \centering
  \includegraphics[width=7.5in]{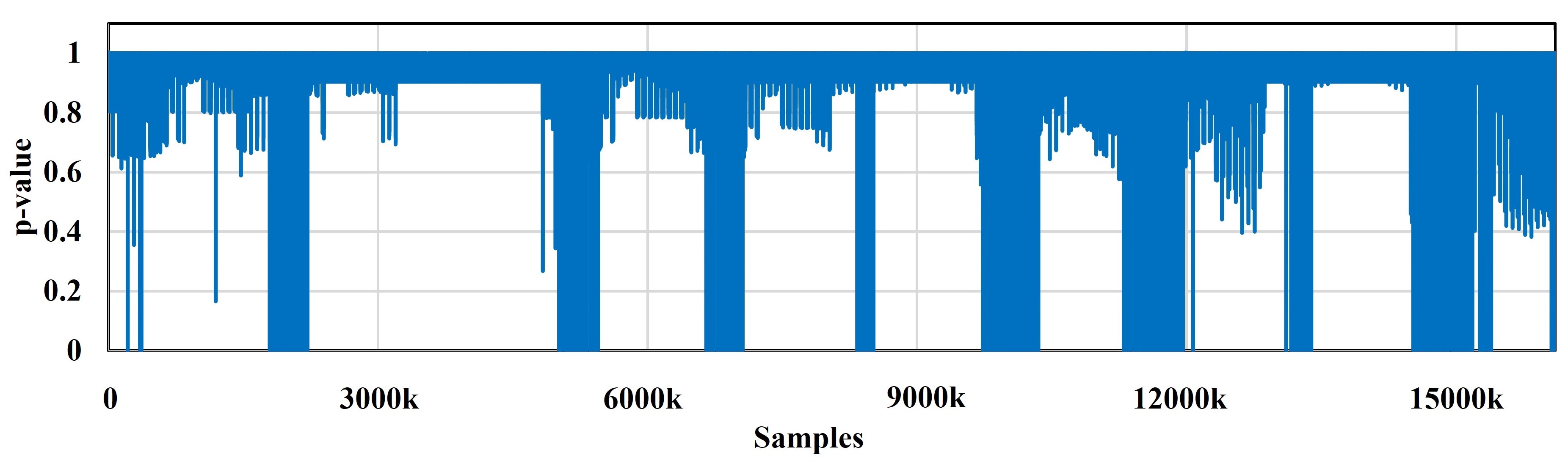}
  \caption{The p-values are computed for 805,000 segments with each segment consisting of 20 samples from the complete signal of 16,100,000 samples (115 fault cases $\times$ 10 fault types $\times$ 14,000 samples).}
  \label{fig:all_p-value}
\end{figure*}

Other fault cases, such as AG fault, ABG fault, and ABCG faults, are also shown in Fig. \ref{fig:other_cases}. The p-values of the complete dataset of 115 cases for 10 fault types are shown in Fig. \ref{fig:all_p-value}. By observing the variation of p-values for our data; the threshold $\alpha$ value is found to be 0.9892. When p-values are less than the threshold value, faults are considered to be detected, and the fault events are recorded until the p-value returns to the threshold value for eight continuous segments. The faults are accurately detected for the entire dataset. The first segments of each fault event are used for the categorization of faults.

\subsubsection{Fault Clustering Results}

\begin{figure}[]
  \centering
  \includegraphics[width=3.4in]{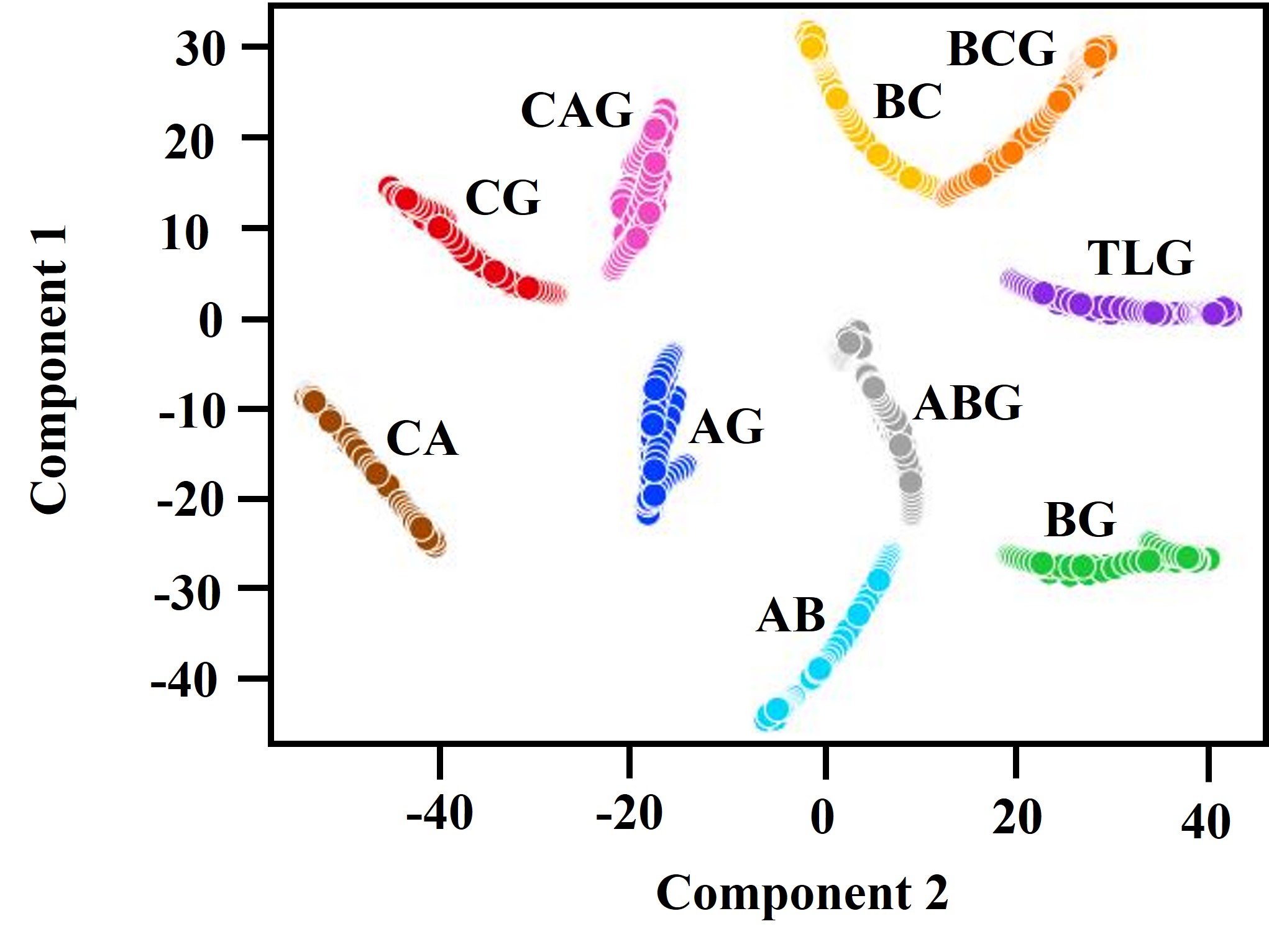}
  \caption{Clusters of different fault types.}
  \label{fig:cluster}
\end{figure}

For the detected fault event, we have considered the first (fault starting) segment and extracted the current and voltage signal values of three phases and 1-D HLLE values. Then, we took the average of each of the seven considered signals. This process yielded 7 feature values for one fault, which are then provided as input to t-SNE for dimensionality reduction. The t-SNE helps to map HD data to a lower-dimensional space where similar data points are represented closer to each other, making it useful for visualizing complex data structures. Then, we performed Gaussian mixture model clustering to identify the underlying distribution of data points and categorize faults based on their types. We simulated 115 fault cases for each fault type, and after their detection, we clustered them as shown in Fig. \ref{fig:cluster}. 

The scores obtained for clustering the faults based on their types are tabulated in Table \ref{tab:scores}; which indicate distinct clusters for all fault types with high accuracy. The clustering scores are used to evaluate the quality and effectiveness of clustering algorithms. These scores provide numerical measures that assess different aspects of clustering results. Generally, $MI$ measures the amount of information shared between true cluster assignments and the clusters produced by an algorithm. $MI$ score is equal to 2.3, indicating strong agreement between the two, making it significant for assessing the overall quality and informativeness of clusters. $AMI$ score of 1.0 signifies that the clustering results are highly meaningful, demonstrating the significance of their agreement with true clusters. The Rand index quantifies the similarity between true and predicted clusters by considering the number of agreements and disagreements in cluster assignments. $RI$ score of 1.0 indicates complete agreement between the two, making it a valuable metric for evaluating clustering performance. $ARI$ score of 1.0 signifies a high level of agreement between true and predicted clusters. A completeness score of 1.0 indicates that the algorithm has successfully captured all data points within their true clusters, which is a significant aspect of clustering quality. A homogeneity score of 1.0 suggests that the algorithm has produced clusters that are internally consistent in terms of class membership, emphasizing the significance of cluster purity. The high scores for all these metrics signify that the clustering results are in strong agreement with the true cluster assignments. This indicates the high quality and reliability of the clustering algorithm making it a valuable tool for decision-making.

\begin{table}[t]
\caption{Clustering Scores.}
\centering
\begin{tabular}{lc}
\toprule
Metric & Score \\                      
\midrule
Mutual Info           & 2.3            \\
Adjusted Mututal Info     & 1.0           \\
Rand     & 1.0           \\
Adjusted Rand              & 1.0            \\
Completeness     & 1.0           \\
Homogeneity     & 1.0           \\
\bottomrule
\end{tabular}
\label{tab:scores}
\end{table}

\subsection{Application for Online Fault Detection}
The proposed methodology is apt for online fault detection and clustering. For real-time streaming data, segments of the data are to be created for  HLLE transformation and use of the Mann-Whitney U test with the comparing reference segment to obtain the p-value for each segment. Then by selecting a suitable threshold value, fault events will be detected. Further, as new data is streamed in, for clustering the type of fault events, the t-SNE embedding is updated incrementally to reflect the evolving data distribution, and visualizations help monitor cluster changes over time. Then, for clustering, we begin by constructing an initial GMM from a portion of the data. As new data arrives, the model is updated using the EM algorithm, and data points are assigned to clusters based on their probabilities. Cluster properties are adjusted as well, and clusters are dynamically merged or split to adapt to changing data patterns. Online clustering with the proposed methods would be useful for analyzing streaming or large datasets while maintaining the flexibility to adapt to data shifts.

\section{Conclusion}
\label{sec:conclusion}

In this work, we introduced a novel methodology that uses Hessian locally linear embedding with Mann-Whitney U statistical test for efficient electrical fault detection and t-distributed stochastic neighbor embedding with a Gaussian mixture model for fault categorization. A large fault dataset is simulated with varied fault resistances and locations for each fault type. The proposed method showed consistency in detecting each fault and also accurately clustered the faults into separate clusters for each fault type. The fault clusters are verified with known labels from simulation, resulting in very high accuracy. Thus, the proposed methodology could offer practical benefits in terms of enhancing fault detection accuracy, reducing downtime and improving system reliability. Future research could focus on testing the proposed methodology for a wide range of events and exploring its applicability in other domains beyond electrical systems. Additionally, integrating real-time data and adaptive learning algorithms could further enhance its capabilities.

\bibliographystyle{ieeetr}
\bibliography{main}
\end{document}